\PassOptionsToPackage{pdfpagelabels=false}{hyperref}
\documentclass[useAMS,usenatbib]{mnras}

\usepackage[pdftex]{graphicx}
\usepackage{color}
\usepackage{amsmath}  
\usepackage{amssymb}  
\usepackage{upgreek}  
\usepackage{afterpage}  
\usepackage{times}  
\usepackage{paralist}  
\usepackage{pbox}  
\usepackage{booktabs}  
\usepackage[flushleft]{threeparttable}  
\usepackage{mathtools}  

\usepackage{grffile}

\newcommand{\Msun}{{\rm M}_{\odot}}

\title[Optical morphology in IllustrisTNG]{The optical morphologies of galaxies in the IllustrisTNG simulation: \\ a comparison to Pan-STARRS observations}

\author[V. Rodriguez-Gomez et al.]
{
	\parbox{18cm}{
    Vicente Rodriguez-Gomez,$^{1}$\thanks{E-mail: vrg@jhu.edu}
		Gregory F. Snyder,$^{2}$
		Jennifer M. Lotz,$^{2}$
		Dylan Nelson,$^{3}$ \\
		Annalisa Pillepich,$^{4}$
		Volker Springel,$^{3}$
		Shy Genel,$^{5,6}$
		Rainer Weinberger,$^{7}$ \\
		Sandro Tacchella,$^{7}$
		R\"{u}diger Pakmor,$^{8}$
		Paul Torrey,$^{9}$
		Federico Marinacci,$^{10,7}$ \\
		Mark Vogelsberger,$^{10}$
		Lars Hernquist$^{7}$
		and David A. Thilker$^{1}$
	}
	\vspace{0.3cm} \\ 
	$^{1}$ Department of Physics \& Astronomy, Johns Hopkins University, 3400 N. Charles Street, Baltimore, MD 21218, USA \\
	$^{2}$ Space Telescope Science Institute, 3700 San Martin Drive, Baltimore, MD 21218, USA \\
    $^{3}$ Max-Planck-Institut f\"{u}r Astrophysik, Karl-Schwarzschild-Stra\ss{}e 1, D-85741 Garching bei M\"{u}nchen, Germany \\
	$^{4}$ Max-Planck-Institut f\"{u}r Astronomie, K\"{o}nigstuhl 17, D-69117 Heidelberg, Germany \\
	$^{5}$ Center for Computational Astrophysics, Flatiron Institute, 162 5th Avenue, New York, NY 10010, USA \\
	$^{6}$ Columbia Astrophysics Laboratory, Columbia University, 550 West 120th Street, New York, NY 10027, USA \\
	$^{7}$ Harvard-Smithsonian Center for Astrophysics, 60 Garden Street, Cambridge, MA 02138, USA \\
	$^{8}$ Heidelberg Institute for Theoretical Studies, Schloss-Wolfsbrunnenweg 35, D-69118 Heidelberg, Germany \\
	$^{9}$ Department of Astronomy, University of Florida, 211 Bryant Space Sciences Center, Gainesville, FL 32611, USA \\
	$^{10}$ Department of Physics, Kavli Institute for Astrophysics and Space Research, Massachusetts Institute of Technology, Cambridge, MA 02139, USA
}

\begin{document}


\maketitle
\begin{abstract}
We have generated synthetic images of $\sim$27,000 galaxies from the IllustrisTNG and the original Illustris hydrodynamic cosmological simulations, designed to match Pan-STARRS observations of $\log_{10}(M_{\ast}/{\rm M}_{\odot}) \approx 9.8$--$11.3$ galaxies at $z \approx 0.05$. Most of our synthetic images were created with the \textsc{skirt} radiative transfer code, including the effects of dust attenuation and scattering, and performing the radiative transfer directly on the Voronoi mesh used by the simulations themselves. We have analysed both our synthetic and real Pan-STARRS images with the newly developed \texttt{statmorph} code, which calculates non-parametric morphological diagnostics -- including the Gini--$M_{20}$ and concentration--asymmetry--smoothness (CAS) statistics -- and performs 2D S\'{e}rsic fits. Overall, we find that the optical morphologies of IllustrisTNG galaxies are in good agreement with observations, and represent a substantial improvement compared to the original Illustris simulation. In particular, the locus of the Gini--$M_{20}$ diagram is consistent with that inferred from observations, while the median trends with stellar mass of all the morphological, size and shape parameters considered in this work lie within the $\sim$1$\sigma$ scatter of the observational trends. However, the IllustrisTNG model has some difficulty with more stringent tests, such as producing a strong morphology--colour relation. This results in a somewhat higher fraction of red discs and blue spheroids compared to observations. Similarly, the morphology--size relation is problematic: while observations show that discs tend to be larger than spheroids at a fixed stellar mass, such a trend is not present in IllustrisTNG.

\end{abstract}


\section{Introduction}\label{sec:intro}
\renewcommand{\thefootnote}{\fnsymbol{footnote}}

\renewcommand{\thefootnote}{\arabic{footnote}}

The past few years have seen substantial progress in our understanding of galaxy formation and evolution from a statistical perspective, both on the observational side with surveys such as the Cosmic Assembly Near-infrared Deep Extragalactic Legacy Survey \citep[CANDELS,][]{Grogin2011} and the Sloan Digital Sky Survey \citep[SDSS,][]{York2000}, and on the theoretical side with large-scale hydrodynamic cosmological simulations such as Illustris \citep{Vogelsberger2014a, Vogelsberger2014, Genel2014a}, EAGLE \citep{Schaye2015, Crain2015}, and Horizon-AGN \citep{Dubois2014}. However, galaxy formation is complex and fully understanding it will require increased synergy between observations and theory.

One way of achieving such increased interaction between theory and observations is by means of so-called `forward modelling' of simulation data. In particular, the generation and subsequent analysis of synthetic images from hydrodynamic simulations is quickly becoming a powerful tool to connect galaxy formation theory with observations \citep[e.g.][]{Lotz2008, Jonsson2010, Snyder2015a, Snyder2015, Bignone2017, Bottrell2017a, Trayford2017}. Nevertheless, the generation of realistic synthetic images is not without challenges of its own. While some aspects of synthetic images are relatively well understood and standardized by now, such as the use of stellar population synthesis models \citep[e.g.][]{Bruzual2003}, other aspects have been a consistent challenge for theorists, such as the detailed modelling of dust absorption and scattering.

Some simple models that estimate the overall, \textit{spatially unresolved} effects of dust on starlight have been found to work well in a variety of situations \citep[e.g.][]{Charlot2000}. However, quantifying the \textit{spatially resolved} effects of dust on galaxy structure is particularly challenging because this requires modelling a very large ensemble of photons as it travels through the interstellar medium (ISM), considering wavelength-dependent absorption and scattering events along the way. This complex problem is usually known as \textit{radiative transfer} in astrophysics, and is closely related to the concept of \textit{ray tracing} in other fields, such as computer science. Thankfully, there are many publicly available codes designed for radiative transfer calculations in astrophysics, such as \textsc{sunrise} \citep{Jonsson2006}, \textsc{hyperion} \citep{Robitaille2011} and \textsc{skirt} \citep{Baes2011, Camps2015}.

In this paper, we create realistic synthetic images from the state-of-the-art IllustrisTNG simulation \citep{Marinacci2018, Naiman2018, Nelson2018, Pillepich2018a, Springel2018} using the radiative transfer code \textsc{skirt} \citep{Baes2011, Camps2015}, designing them to match observations of low-redshift galaxies from the Pan-STARRS $3\uppi$ Steradian Survey \citep{Chambers2016}. Then, we quantify various structural parameters of both the simulated and observed galaxies using the same morphology code, which allows us to make a fair comparison between theory and observations. This constitutes the first systematic study of image-based galaxy morphology from a hydrodynamic cosmological simulation while including the effects of (spatially resolved) dust attenuation and scattering.

This paper is organized as follows. In Section \ref{sec:methodology}, we introduce the simulations used for this work and define the observational and simulated galaxy samples. In Section \ref{sec:synthetic_images}, we explain how synthetic images are created using the radiative transfer code \textsc{skirt}. In Section \ref{sec:optical_morphologies}, we describe the newly developed \texttt{statmorph} code, which we use to measure various image-based morphological diagnostics. Our main results are presented in Section \ref{sec:results}. Finally, we discuss these results and present our conclusions in Section \ref{sec:discussion_and_conclusions}.

\section{Methodology}\label{sec:methodology}

\subsection{The IllustrisTNG simulation suite}\label{subsec:illustrisTNG}

The IllustrisTNG Project \citep{Marinacci2018, Naiman2018, Nelson2018, Pillepich2018a, Springel2018} is a suite of magneto-hydrodynamic cosmological simulations run with the moving-mesh code \textsc{arepo} \citep{Springel2010, Pakmor2011, Pakmor2016}, featuring an updated version of the Illustris galaxy formation model \citep{Vogelsberger2013, Torrey2014} described in \cite{Weinberger2017} and \cite{Pillepich2018}. In this study, we use the highest resolution version of `TNG100' (hereafter referred to as \textit{the} IllustrisTNG simulation), which follows the evolution of $2 \times 1820^3$ resolution elements within a periodic cube measuring $75h^{-1} \approx 110.7$ Mpc per side, which translates to an average mass of the baryonic resolution elements of $1.39 \times 10^6 \, \Msun$. The gravitational softening length of the DM and stellar particles, which essentially sets the spatial scale of the simulation, is $0.5 h^{-1} \approx 0.74$ kpc, which makes it ideal for a comparison to the Pan-STARRS observations described in Section \ref{subsec:obs_sample}.

The IllustrisTNG model was roughly tuned to match the following observables: (i) the global star formation rate (SFR) density at $z=0$--$8$, (ii) the galaxy mass function at $z=0$, (iii) the stellar-to-halo mass relation at $z=0$, (iv) the black hole-to-stellar mass relation at $z=0$, (v) the halo gas fraction at $z=0$, and (vi) galaxy sizes at $z=0$. Note, however, that the model was not tuned to match galaxy morphology, much less using image-based statistics such as the ones presented in this paper.

We also use data from the original Illustris simulation \citep{Vogelsberger2014a, Vogelsberger2014, Genel2014a, Sijacki2015, Nelson2015}, which follows the evolution of $2 \times 1820^3$ resolution elements within a periodic cube measuring $75h^{-1} \approx 106.5$ Mpc per side, with a baryonic mass resolution of $1.26 \times 10^6 \, \Msun$ and a gravitational softening length of $0.5h^{-1} \approx 0.71$ kpc. The main updates to the original Illustris galaxy formation model \citep{Vogelsberger2013, Torrey2014} implemented in IllustrisTNG, besides the inclusion of ideal magneto-hydrodynamics, consist of a new active galactic nucleus (AGN) feedback model that operates at low accretion rates \citep{Weinberger2017} and various modifications to the galactic winds, stellar evolution, and chemical enrichment schemes \citep{Pillepich2018}. In both the IllustrisTNG and original Illustris simulations, galaxies were identified with the \textsc{subfind} halo-finding algorithm \citep{Springel2001, Dolag2009a}.

Unless noted otherwise, the cosmological parameters used throughout this work are $\Omega_{\rm m} = 0.3089$, $\Omega_{\rm b} = 0.0486$, $\Omega_{\Lambda} = 0.6911$, $\sigma_{8} = 0.8159$, $n_{\rm s} = 0.9667$ and $h = 0.6774$, consistent with recent Planck measurements \citep{Ade2016}.\footnote{However, when showing results from the original Illustris simulation, we use nine-year \textit{Wilkinson Microwave Anisotropy Probe} observations \citep{Hinshaw2013}: $\Omega_{\rm m} = 0.2726$, $\Omega_{\Lambda} = 0.7274$, $\Omega_{\rm b} = 0.0456$, $\sigma_8 = 0.809$, $n_{\rm s} = 0.963$, and $h = 0.704$.}

\subsection{The observational galaxy sample}\label{subsec:obs_sample}

\begin{figure}
  \centering
	\includegraphics[width=8cm]{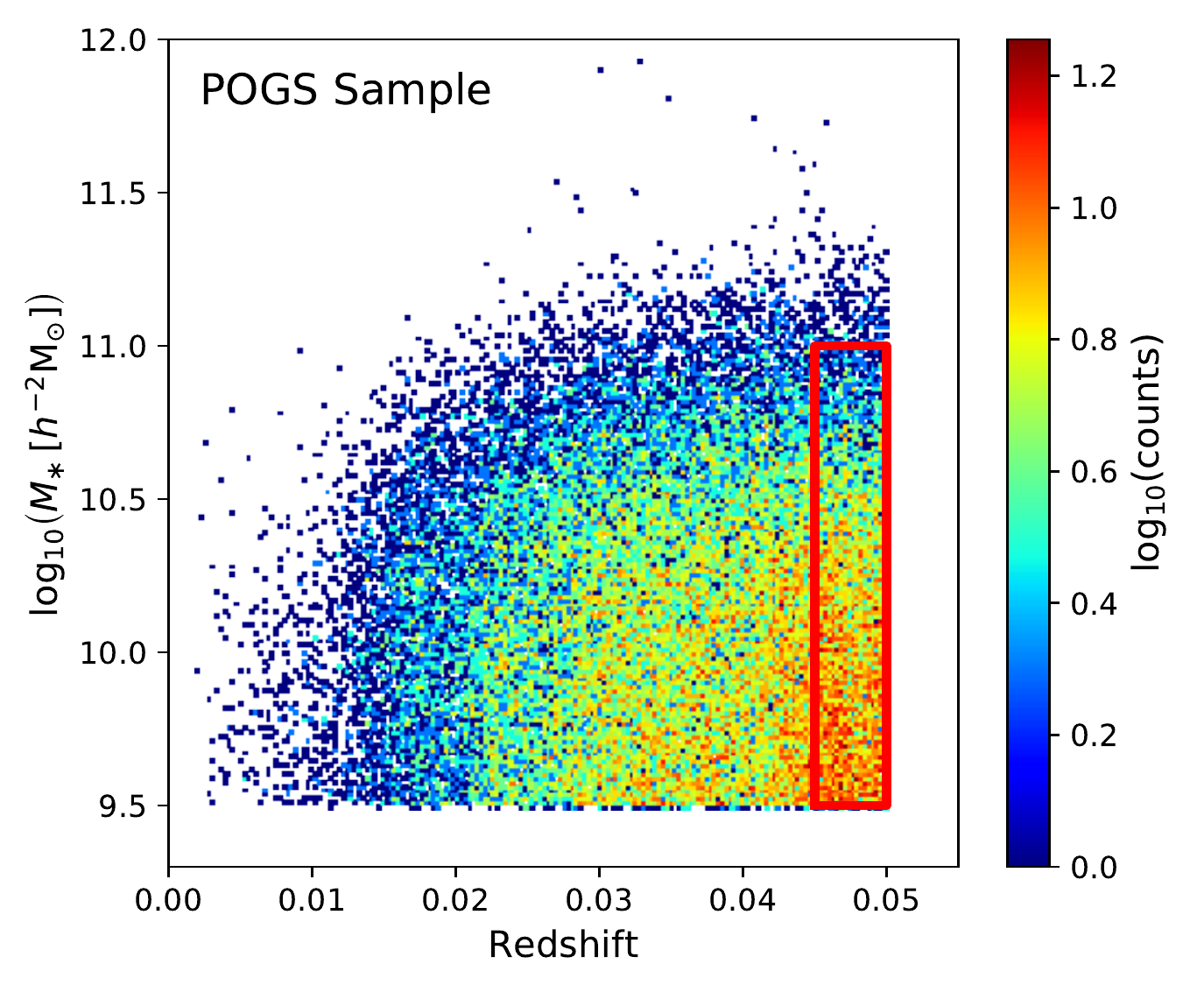}
	\caption{The distribution of stellar masses and redshifts of 56,614 galaxies from the POGS sample. The red rectangle indicates the redshift and stellar mass range considered in our study. The colour scale indicates the number of galaxies per 2D bin.}
	\label{fig:pogs_sample}
\end{figure}

We consider galaxies observed with the Panoramic Survey Telescope and Rapid Response System (Pan-STARRS), specifically the first completed component, Pan-STARRS1 (PS1), which consists of a 1.8 m telescope equipped with a 1.4 gigapixel camera, producing images with a pixel scale of 0.258 arcsec pixel$^{-1}$. The PS1 instrument has been used in a number of distinct sky surveys, in particular the $3\uppi$ Steradian Survey, which mapped the northern hemisphere using five photometric filters (PS1 $g,r,i,z,y$) at an angular resolution of $\sim$1 arcsec and with a $5\sigma$ point source limiting sensitivity of up to $\sim$23 mag \citep{Chambers2016}. The final, stacked image products have a pixel scale of 0.25 arcsec pixel$^{-1}$, slightly smaller than the native pixel scale, as a result of the \textsc{warp} pixel processing stage \citep{Waters2016}.

A catalogue of galaxies from the $3 \uppi$ Survey can be found in the PS1 Optical Galaxy Survey (POGS), an ongoing project that aims to create a value-added, multiwavelength atlas of galaxies in the nearby Universe (\citealt{Vinsen2013}; Thilker et al., in preparation). For each galaxy included in the POGS catalogue, we use its $i$-band stacked image (along with its associated segmentation map, stellar mask, weight map, and point spread function, PSF) for the morphological analysis presented in Section \ref{sec:results}.

We obtain the stellar masses and redshifts of POGS galaxies using the NASA Sloan Atlas (NSA),\footnote{http://www.nsatlas.org/data} keeping only galaxies that are successfully matched: those with angular separations between POGS and NSA coordinates smaller than 2.5 arcsec. From an initial sample of 62,061 POGS galaxies, our coordinate matching criterion results in a sample of 56,614 objects with NSA counterparts, which is presented in Fig. \ref{fig:pogs_sample}. This figure shows that, by construction, POGS galaxies have stellar masses $M_{\ast} > 10^{9.5} h^{-2} \, \Msun$ and redshifts $z < 0.05$.

In order to make a straightforward comparison to a single snapshot from the IllustrisTNG and original Illustris simulations, we only consider galaxies in the redshift interval $z = 0.045$--$0.05$. Within this redshift range, we find that the completeness of the POGS sample (relative to the NSA) is high, but falls below 90 per cent at $M_{\ast} \gtrsim 10^{11} h^{-2} \, \Msun$. Therefore, we only consider POGS galaxies with stellar masses in the range $M_{\ast} = 10^{9.5}$--$10^{11} h^{-2} \, \Msun$, or $M_{\ast} \approx 10^{9.8}$--$10^{11.3} \, \Msun$. We also note that the spatial resolution of PS1 observations for this redshift range is $\sim$1 kpc, which is ideal for a comparison with the simulations described in Section \ref{subsec:illustrisTNG}. Our final POGS sample, which consists of 10,829 galaxies, is indicated with a red rectangle in Fig. \ref{fig:pogs_sample}.

\subsection{The simulated galaxy sample}\label{subsec:sim_sample}

On the simulation side, we consider a single simulation snapshot at $z = 0.0485$ (snapshot 95 in IllustrisTNG, or snapshot 131 in Illustris original), consistent with the observational redshift range mentioned in Section \ref{subsec:obs_sample}. Within this snapshot, we consider all simulated galaxies with $M_{\ast} > 10^{9.5} \, \Msun$, a stellar mass range that encompasses that of the POGS sample. Such galaxies are typically resolved with more than $\sim$2,250 ($\sim$2,500) stellar particles in the IllustrisTNG (Illustris original) simulation. The resulting mock galaxy catalogue, which includes both central and satellite galaxies, consists of 12,470 (14,721) galaxies in IllustrisTNG (Illustris original).

\section{Generating synthetic images}\label{sec:synthetic_images}

\begin{figure}
  \centering
	\includegraphics[width=8cm]{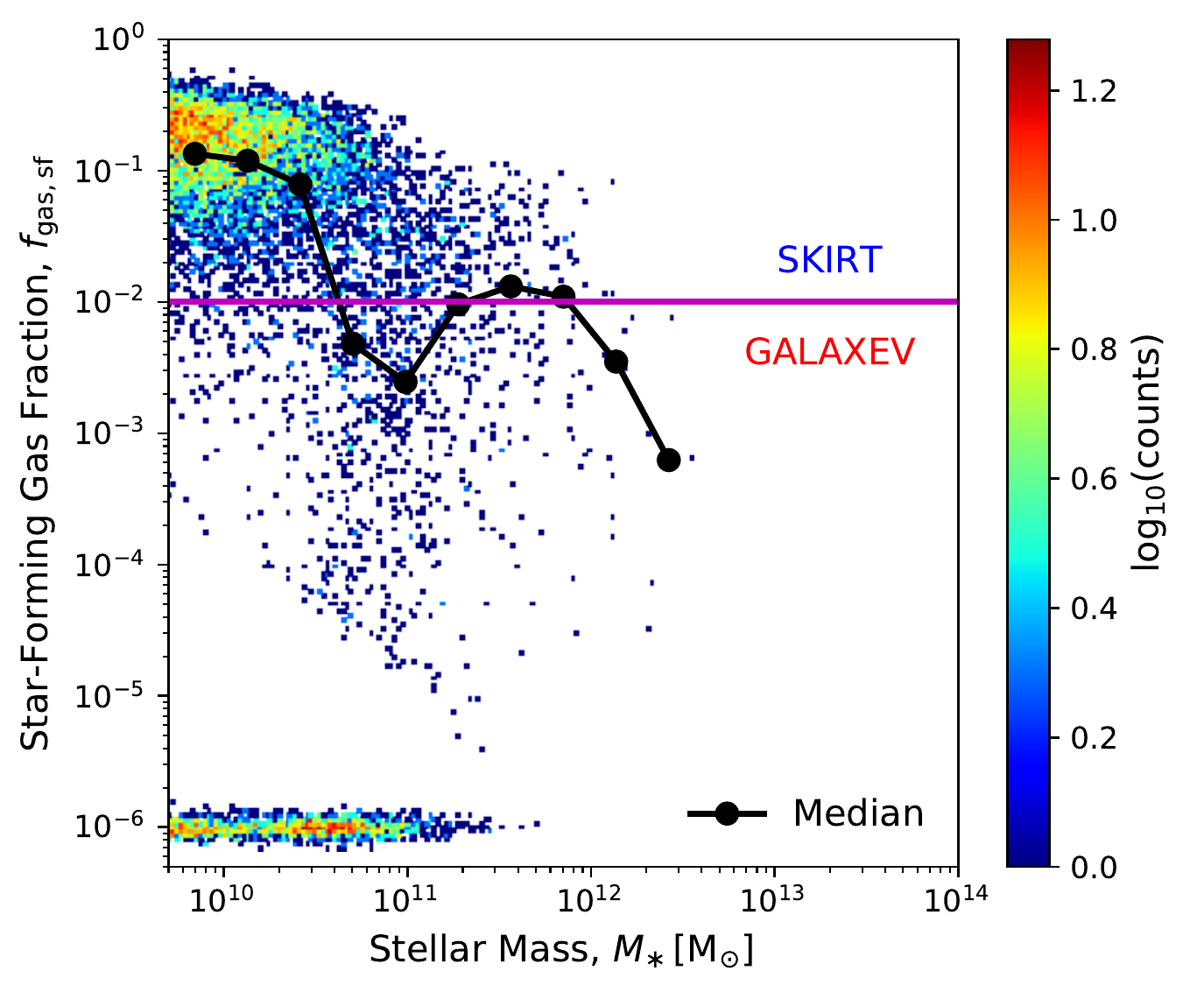}
	\caption{The fraction of star-forming gas as a function of stellar mass for all galaxies in snapshot 95 from IllustrisTNG, which corresponds to $z = 0.0485$. The 2D histogram shows the overall galaxy distribution, while the black line shows the median trend. Synthetic images of galaxies above the magenta line are created with our \textsc{skirt} pipeline (Section \ref{subsec:skirt_pipeline}), while galaxies below the line are processed with our \textsc{galaxev} pipeline (Section \ref{subsec:galaxev_pipeline}). Galaxies with zero star-forming gas are shown at $f_{\rm gas, sf} \approx 10^{-6}$.}
	\label{fig:fgas_vs_mstar}
\end{figure}

In order to generate reasonably realistic synthetic images from our simulations, we not only predict the light distribution produced by stellar populations and star-forming regions, but also model the effects of dust attenuation and scattering as the starlight traverses the ISM. This requires radiative transfer calculations, which can be very computationally expensive for large images (especially above a thousand pixels on each dimension), and not very worthwhile for galaxies with negligible amounts of star-forming gas, which we use as a proxy for dust (see Section \ref{subsubsec:dust_modelling}). Therefore, in order to reduce the overall computational cost of our synthetic images, we have developed two synthetic image generation pipelines, with and without dust radiative transfer, which we refer to as our `\textsc{skirt} pipeline' and `\textsc{galaxev} pipeline', respectively. As their names suggest, our pipelines use the \textsc{skirt} radiative transfer code \citep{Baes2011, Camps2015} and the \textsc{galaxev} stellar population synthesis code \citep{Bruzual2003}. We process galaxies with negligible fractions of star-forming gas using our inexpensive \textsc{galaxev} pipeline, while the rest are treated with full radiative transfer using our \textsc{skirt} pipeline. The threshold is defined at a star-forming gas fraction $f_{\rm gas, sf} = 1/100$ (with respect to the total baryonic mass), as illustrated in Fig. \ref{fig:fgas_vs_mstar}. This is approximately the value at which we find that dust effects begin to have a noticeable impact on galaxy morphology.

Throughout this section, we express some distances in comoving coordinates (which is how distances are handled internally by the \textsc{arepo} code), which we indicate with a `c' prefix. In particular, we refer to comoving kpc as `ckpc'.

\subsection{General details}\label{subsec:general_details}

Several aspects of the synthetic images are the same for both our \textsc{skirt} and \textsc{galaxev} pipelines. Each galaxy is observed from a viewing angle perpendicular to the $xy$-plane of the simulation volume, which essentially corresponds to a random orientation for each galaxy. The field of view of each image is equal to 15 times the (3D) stellar half-mass radius of the corresponding galaxy in IllustrisTNG, and 10 times in Illustris original. These values were chosen empirically, so that the final image products are large enough for the morphological measurements described in Section \ref{sec:optical_morphologies}, after adding an adequate amount of background noise (Section \ref{subsubsec:noise}). The number of pixels is chosen so that the resulting pixel scale matches that of the observations (0.25 arcsec pixel$^{-1}$, which corresponds to $0.174 \, h^{-1}$ ckpc pixel$^{-1}$ at $z = 0.0485$).

For each stellar particle, we define an adaptive smoothing scale equal to the (3D) distance to the $N$th nearest stellar particle. This approach was described in \cite{Torrey2015}, although we use $N=32$ instead of $N=16$. We find that our morphological measurements are not particularly sensitive to the number of neighbours used, at least for choices ranging from $N=4$ to $N=64$.

Unless noted otherwise, all smoothing calculations in this section are carried out through a convolution with a `standard' smoothed particle hydrodynamics (SPH) spline kernel \citep{Monaghan1985, Hernquist1989, Monaghan1992} redefined over the interval $\left[0, h_{\rm sml}\right]$ \citep{Springel2001a}, where $h_{\rm sml}$ is some appropriately chosen smoothing length:

\begin{flalign}
W(r, h_{\rm sml}) = \frac{2^{\nu} \sigma}{h_{\rm sml}^{\nu}} \left\{\begin{matrix*}[l]
1 - 6 q^2 + 6 q^3 & {\rm if} & 0 \leq q \leq \displaystyle\frac{1}{2},  \\
2(1-q)^3 & {\rm if} & \displaystyle\frac{1}{2} < q \leq 1, \\ 
0 & {\rm if} & q > 1, 
\end{matrix*}\right. &&
\label{eq:spline_kernel}
\end{flalign}
where $q \equiv r / h_{\rm sml}$, $r$ is the radial distance, $\nu$ is the number of dimensions, and $\sigma$ is a normalization constant with the values
\begin{flalign*}
\frac{2}{3}, \hspace{12pt} \frac{10}{7\uppi}, \hspace{12pt} \frac{1}{\uppi} &&
\end{flalign*}
in one, two, and three dimensions, respectively.

\subsection{\textsc{skirt} pipeline}\label{subsec:skirt_pipeline}

In this section, we describe our \textsc{skirt} synthetic image generation pipeline, based on the radiative transfer code \textsc{skirt} \citep{Baes2011, Camps2015}, which we use to produce synthetic images including the effects of dust attenuation and scattering.

\subsubsection{Stellar sources}\label{subsubsec:stellar_sources}

Each stellar particle in the simulation represents a coeval stellar population. The spectral energy distributions (SEDs) of stellar particles older than 10 Myr are modelled with the \textsc{galaxev} population synthesis code \citep{Bruzual2003}, as implemented internally in \textsc{skirt}. The user must simply provide the initial mass (i.e. neglecting mass loss due to stellar evolution), metallicity, and stellar age of each stellar particle. More details about \textsc{galaxev} can be found in Section \ref{subsec:galaxev_pipeline}.

On the other hand, stellar populations younger than 10 Myr are essentially treated as starbursting regions, and their SEDs are modelled with the \textsc{mappings-iii} photoionization code \citep{Groves2008}, which includes emission from H \textsc{ii} regions and their surrounding photodissociation regions (PDRs), as well as absorption by gas and dust in the `birth clouds' associated with such star-forming regions. The \textsc{mappings-iii} models require the following five parameters for each star-forming region: (i) the SFR, assumed to be constant over the last 10 Myr, (ii) the metallicity, (iii) the compactness parameter $\mathcal{C}$, (iv) the ISM pressure $P_{0}$, and (v) the cloud covering fraction, $f_{\rm PDR}$. For every young stellar particle, we assume that the SFR is given by its initial mass divided by 10 Myr, and use its nominal metallicity (inherited from its parent gas cell). Since changes in the compactness parameter $\mathcal{C}$ or in the ISM pressure $P_0$ only have a noticeable effect on the far infrared regime of the SEDs, which we do not consider in this work, we simply use `typical' fixed values of $\log_{10} \mathcal{C} = 5$ and $\log_{10} [(P_{0}/k_{\rm B}) / {\rm cm^{-3} \, K}] = 5$ \citep{Groves2008}. Following \cite{Jonsson2010}, we adopt a value of $f_{\rm PDR} = 0.2$ for the cloud covering fraction.

We note that, due to the coarse sampling of star formation in cosmological simulations, the spatial distribution of young stellar populations becomes sensitive to stochastic noise \citep[see][for a discussion]{Trayford2015, Nelson2018}. This motivated \cite{Camps2016} and \cite{Trayford2017}, using the \textsc{eagle} cosmological simulation \citep{Schaye2015, Crain2015}, to implement a post-processing resampling technique to model the expected frequency of young stellar populations at a higher resolution. For simplicity, we do not resample the young stellar populations, and defer a detailed quantification of the impact of such resampling techniques to future work.

Finally, as mentioned in Section \ref{subsec:general_details}, every stellar particle is associated with a smoothing scale equal to the distance to its 32nd nearest neighbour. This scale is used internally by \textsc{skirt} to smooth the stellar mass density distribution, using the 3D version of the SPH kernel introduced in equation (\ref{eq:spline_kernel}).

\subsubsection{Dust modelling}\label{subsubsec:dust_modelling}

One remarkable feature of \textsc{skirt} is that it provides the option of performing the radiative transfer calculations directly on a 3D Voronoi mesh \citep{Camps2013}, which is particularly well suited for hydrodynamic simulations run with \textsc{arepo} \citep{Springel2010}. This means that we can reconstruct the gas (and metal) density distributions in a self-consistent fashion, i.e. exactly as implemented in the hydrodynamic solver (except for the cell gradients) in order to dynamically evolve the system. Therefore, for each simulated galaxy, we consider a cubical region around it with the same dimensions as the field of view (Section \ref{subsec:general_details}). The positions of the gas cells -- which are actually mesh-generating points -- are used by \textsc{skirt} to uniquely reconstruct the Voronoi mesh inside this volume.\footnote{Behind the scenes, this is done using the \textsc{voro++} open source library for computing Voronoi tessellations \citep{Rycroft2009}.} This, along with the recorded density value for each gas cell, completely determines a gas density distribution for each galaxy.

In order to convert the gas density distribution into a dust density distribution, we assume that the diffuse dust content of each galaxy is traced by the star-forming gas.\footnote{The simulations used in this work do not track dust explicitly, although impressive progress has been made in this direction \citep{McKinnon2016, McKinnon2017, McKinnon2018}.} The latter consists of gas cells above a given density threshold ($n_{\rm H} \sim 0.1 \, {\rm cm}^{-3}$), which are stochastically converted into stellar particles \citep[for details, see][]{Springel2003, Vogelsberger2013, Pillepich2018}. The dust densities of gas cells that are not star-forming, as well as gas cells that are not part of the galaxy in question, are set to zero. For the remaining gas cells, we assume a constant dust-to-metal mass ratio of 0.3 \citep{Camps2016}. This completely determines the overall dust density distribution. Finally, \textsc{skirt} offers several options to model the dust composition. Following \cite{Camps2016} and \cite{Trayford2017}, we choose the multicomponent dust mix of \cite{Zubko2004} including graphite grains, silicate grains, and polycyclic aromatic hydrocarbons (PAHs).

We note that the diffuse dust density distribution described in this section is separate from the dust content in birth clouds, which is already included in the \textsc{mappings-iii} models for star-forming regions (see Section \ref{subsubsec:stellar_sources}) before any radiative transfer calculations are done. Since birth clouds have a finite lifetime ($\sim$10 Myr), starlight from older stellar populations is only expected to be absorbed and scattered by diffuse dust in the ambient ISM, which we model using \textsc{skirt}.

\subsubsection{Running \textsc{skirt}}\label{subsubsec:running_skirt}

\textsc{skirt} is a Monte Carlo radiative transfer code, which means that the radiation field is represented as a discrete, but very large, number of `photon packages' (also referred to as `rays'). Typical values for the number of rays (per wavelength) range from $10^{6}$ to $10^{8}$. However, since the sizes of our synthetic images are proportional to the sizes of the galaxies (see Section \ref{subsec:general_details}), choosing any fixed number of rays for the entire galaxy sample would not be optimal. Instead, we define a constant `ray density' of 50 rays pixel$^{-1}$, which we find to be more than sufficient to carry out the morphological measurements described in Section \ref{sec:optical_morphologies}. Similar values for the ray density have been used implicitly in other recent studies \citep[e.g.][]{Snyder2015, Trayford2017}.

Our wavelength grid consists of 50 logarithmically spaced wavelengths between 0.35 and 0.95 $\upmu$m, an interval that encloses the rest-frame and observer-frame PS1 $g,r,i,z$ bands. Although this spectral resolution is not fine enough to adequately resolve some of the narrow emission lines included in the \textsc{mappings-iii} models (Section \ref{subsubsec:stellar_sources}), we find that it yields converged results for the purposes of this paper, such that doubling the resolution of the wavelength grid does not change any of our measurements appreciably. Note that since we are only interested in optical and near-infrared frequencies, thermal dust emission is not included in our analysis.

\subsubsection{Broad-band integration}

\begin{figure*}
  \centering
  \vbox{
	\includegraphics[width=17.5cm]{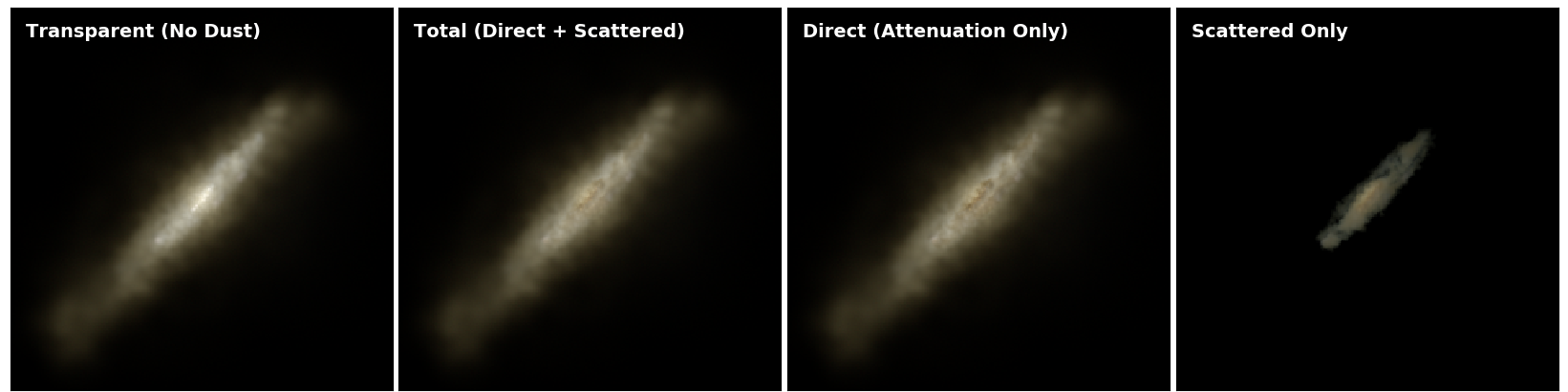}
	\includegraphics[width=17.5cm]{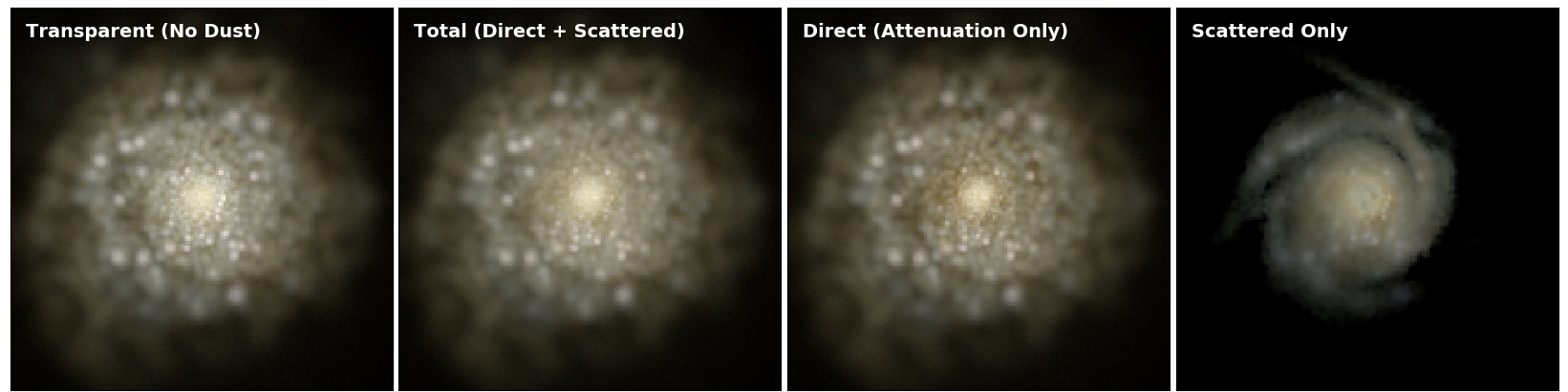}
	\includegraphics[width=17.5cm]{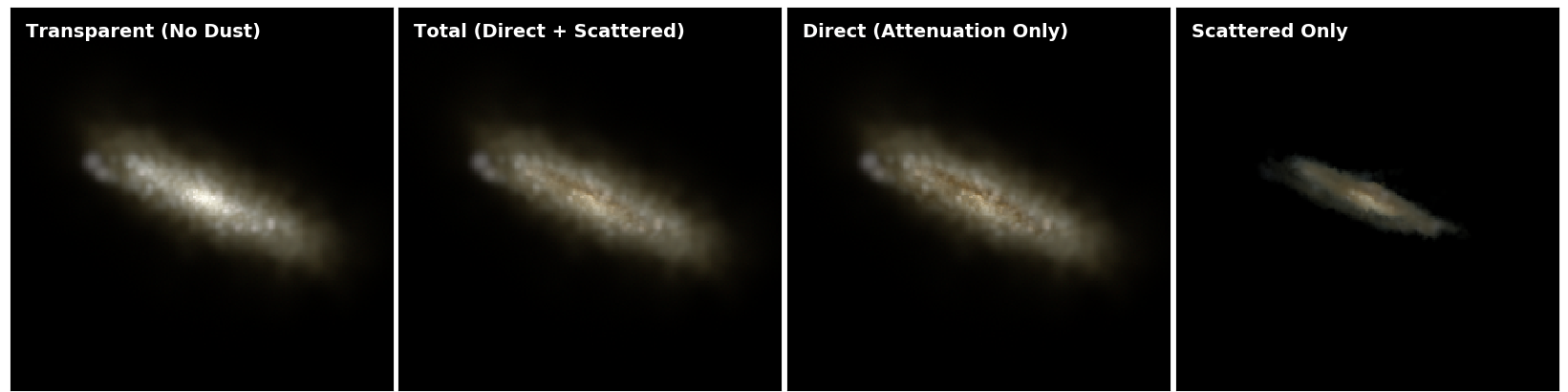}	
  }
	\caption{Idealized synthetic images (using the PS1 $g,r,i$ filters) of randomly selected $M_{\ast} \approx 10^{10.5} \, \Msun$ galaxies from the IllustrisTNG simulation at $z=0.0485$, showing the different light components. From left to right, the different columns show (a) the light distribution that would be observed without any dust effects, (b) the light distribution obtained with the full dust modelling, (c) the contribution of rays that have not been scattered, and (d) the contribution from rays that have been scattered at least once. The field of view in each panel corresponds to 10 stellar half-mass radii. Note that the inclusion of dust tends to reduce the light concentration near the galactic centre.}
	\label{fig:components}
\end{figure*}

The main data product generated by \textsc{skirt} is a 3D data cube for each galaxy, consisting of a full rest-frame SED for each pixel. We then assume that the source is located at $z=0.0485$ and generate the data cube that would be measured by a local observer, taking cosmological effects into account (e.g. surface brightness dimming). Each SED is then multiplied by each of the PS1 $g,r,i,z$ filter curves \citep{Tonry2012} and integrated over the full wavelength range. This procedure results in \textit{idealized} synthetic images for the PS1 $g,r,i,z$ filters as would be hypothetically observed by an instrument with a point-like PSF (i.e. a Dirac delta function) and an infinite signal-to-noise (S/N) ratio. Fig. \ref{fig:components} shows examples of such idealized synthetic images for randomly selected IllustrisTNG galaxies, showing the light distribution that would be obtained without and with dust modelling (first two columns), and separating the dust-modelled light distribution from the second column into non-scattered and scattered rays (third and fourth columns, respectively).

\subsection{\textsc{galaxev} pipeline}\label{subsec:galaxev_pipeline}

Since radiative transfer calculations are computationally intensive, and not particularly worthwhile for galaxies with little dust content, we have also developed a synthetic image generation pipeline based solely on the stellar population synthesis code \textsc{galaxev} \citep{Bruzual2003}. In the absence of dust, this pipeline produces results that are effectively indistinguishable to those of our \textsc{skirt} pipeline, and yet is at least two orders of magnitude faster due to the absence of radiative transfer calculations.

\subsubsection{Stellar sources}\label{subsubsec:stellar_sources_galaxev}

In order to model the spectra of stellar particles in the simulation, we consider the `default' simple stellar population (SSP) models included in \textsc{galaxev}, which were computed using the `Padova 1994' evolutionary tracks and a \cite{Chabrier2003} initial mass function (IMF). These models provide the \textit{rest-frame} luminosity per unit wavelength of an SSP, $L_{\lambda} (\lambda, t, Z)$, as a function of wavelength $\lambda$, age $t$, and metallicity $Z$ (the luminosity is normalized to a mass of $1 \, \Msun$). In practice, this quantity is provided as a `data cube' sampled at 221 unequally spaced SSP ages between 0 and 20 Gyr, seven metallicities between $10^{-4}$ and $0.1$,\footnote{We use a relatively recent release of \textsc{galaxev} that extends the upper limit of the metallicity range from 0.05 to 0.1.} and 1221 wavelengths between $91 \, {\rm \AA}$ and $160 \, \upmu$m.\footnote{We use the lower-resolution BaSeL 3.1 spectral library instead of the higher-resolution STELIB measurements, mostly for consistency with the \textsc{skirt} implementation of \textsc{galaxev}. We find that the mean variation in the PS1 $g, r, i, z$ integrated fluxes as a result of using different spectral libraries is below a thousandth of a magnitude.} Since the spectra of the stellar sources in the \textsc{skirt} pipeline were also calculated using \textsc{galaxev} (Section \ref{subsubsec:stellar_sources}), the only difference between the two pipelines at this stage lies in the modelling of the young stellar populations ($< 10$ Myr), which is done using \textsc{mappings-iii} models (instead of \textsc{galaxev}) in the \textsc{skirt} pipeline. However, such young stellar populations are (by construction) practically absent in those gas-depleted galaxies where the \textsc{galaxev} pipeline replaces the \textsc{skirt} pipeline (see Fig. \ref{fig:fgas_vs_mstar}).

Optionally, a simple \cite{Charlot2000} model can be implemented at this stage in order to estimate the effects of an unresolved dust distribution \citep[e.g.][]{Torrey2015, Trayford2015, Nelson2018}. According to this prescription, the luminosity per unit wavelength of each SSP is attenuated as $L_{\lambda}^{\rm CF00} = L_{\lambda} \exp (-\tau_{\lambda})$, where

\begin{flalign}
\tau_{\lambda} = \left\{\begin{matrix*}[l]
\tau_{1} (\lambda / 5500 \, {\rm \AA})^{\delta} & {\rm if} & t \leq t_{\rm BC},  \\
\tau_{2} (\lambda / 5500  \, {\rm \AA})^{\delta} & {\rm if} & t > t_{\rm BC}. \\ 
\end{matrix*}\right. &&
\end{flalign}

\noindent For simplicity, we adopt the recommended values for all parameters: $\tau_{1} = 1$, $\tau_{2} = 0.3$, $\delta = -0.7$ and $t_{\rm BC} = 10 \, {\rm Myr}$. We emphasize, however, that the inclusion of a \cite{Charlot2000} model in our \textsc{galaxev} pipeline is optional, and in fact we only use it in Section \ref{sec:results} for comparison purposes.

In either case, once the \textit{rest-frame} luminosity per unit wavelength of a given SSP is known, we proceed to calculate the \textit{observer-frame} flux per unit wavelength, taking cosmological effects into account, assuming that each SSP is located at $z = 0.0485$. Finally, given the observer-frame flux per unit wavelength of any SSP, we use appropriate broad-band filter response functions \citep{Tonry2012} to obtain the integrated, apparent $g,r,i,z$ magnitudes that would be measured by the observer. For every broad-band filter, the magnitudes are stored in a `table' (sampled at the same age and metallicity values as the original \textsc{galaxev} data cube) that can be readily interpolated and normalized in order to obtain the apparent $g,r,i,z$ magnitudes of any stellar particle, given its initial mass, age, and metallicity.\footnote{We clip any metallicity values that fall outside the range covered by the \textsc{galaxev} models.}

\subsubsection{Adaptive smoothing}

For each galaxy, an image is created by adding the flux contributions of all stellar particles, assuming that each of them has a spatial distribution given by the \textit{2D} version of the spline kernel given by eq. (\ref{eq:spline_kernel}). The adaptive smoothing length, pixel scale and field of view are the same as those used in our \textsc{skirt} pipeline, and have already been described in Section \ref{subsec:general_details}. This yields \textit{idealized} synthetic images that, in the absence of dust, are equivalent to those produced by our \textsc{skirt} pipeline (Section \ref{subsec:skirt_pipeline}), but are very inexpensive to make.

\subsection{Post-processing}\label{subsec:postprocessing}

As described in the previous sections, both the \textsc{skirt} and \textsc{galaxev} pipelines produce \textit{idealized} synthetic images for the PS1 $g,r,i,z$ broad-band filters. Since the filter curves from \cite{Tonry2012} are given as capture cross-sections with units of ${\rm m}^2 \, e^{-} \, {\rm photon}^{-1}$, the units of the resulting wavelength-integrated images are $e^{-} \, {\rm s}^{-1} \, {\rm pixel}^{-1}$ (electron counts per second per pixel). However, before performing any morphological measurements on these images, we must carry out the post-processing stages described next.

\subsubsection{Convolving with a PSF}\label{subsubsec:psf}

Astronomical images are affected by telescope optics and, in the case of ground-based instruments, by atmospheric noise. These factors can be modelled in our synthetic images by convolving with an appropriate PSF, which for simplicity we choose to be an azimuthally symmetric Gaussian function. We choose the full width at half-maximum (FWHM) of the PSF so that it matches the median seeing of the PS1 instrument, which is 1.11 arcsec (1.31 arcsec) in the $i$ band ($g$ band).

\subsubsection{Modelling noise: $\sigma_{\rm sky}$ and $G$}\label{subsubsec:noise}

Including sky background noise is important not just for the sake of realism, but also because most of the morphological measurements described in Section \ref{sec:optical_morphologies}, as well as the segmentation image described in Section \ref{subsubsec:segmap}, are built upon the notion that the `galaxy' pixels are surrounded by `sky' pixels with random values drawn from some probability distribution. We model sky background noise as a Gaussian random variable that is added to each pixel flux value, with a standard deviation $\sigma_{\rm sky}$ that is constant across the entire image. While $\sigma_{\rm sky}$ could be redefined for each individual image based on a `target' S/N, this would be somewhat inconsistent with data from most galaxy surveys, which by design do not change the exposure time of individual observations arbitrarily. Therefore, we assume the same value of $\sigma_{\rm sky}$ for all of our simulated galaxies.

We find that typical values for the least noisy quarter of POGS galaxies are $\sigma_{\rm sky} = 1/12$ and $1/15 \; e^{-} \, {\rm s}^{-1} \, {\rm pixel}^{-1}$ in the $i$ band and $g$ band, respectively. However, we find that directly applying this level of background noise to our synthetic images results in a somewhat large fraction (up to $\sim$30 per cent) of `bad' measurements (see Section \ref{sec:optical_morphologies}), especially in the $g$ band, at low stellar masses, and for galaxies from the original Illustris simulation, which tend to have lower surface brightness (see also \citealt{Bottrell2017a}, who encountered similar issues while analysing the Illustris simulation). Since it would be instructive to include these galaxies in the analysis, at the cost of sacrificing some realism in the amount of background noise applied to the synthetic images, we choose somewhat lower values for $\sigma_{\rm sky}$. For IllustrisTNG, we adopt values of $\sigma_{\rm sky} = 1/15$ and $1/25 \; e^{-} \, {\rm s}^{-1} \, {\rm pixel}^{-1}$ in the $i$ band and $g$ band, respectively, and half of those values for the original Illustris simulation.

Besides an adequate amount of background noise, some of the morphological measurements presented in Section \ref{sec:optical_morphologies} (namely, the S/N ratio per pixel and the S\'{e}rsic fit parameters) also require a `weight map' that captures the uncertainty in each pixel value. This includes contributions from both the random sky background noise and from Poisson statistics in the number of electron counts measured by the charge-coupled device (CCD) camera, as well as other sources of error (e.g. the `read noise', which for simplicity we will consider to be included in the background noise). While such a weight map could be constructed for each synthetic image and used as input for the morphology code, a simpler approach is to define a `gain' factor that is used internally by the code to generate such a weight map, according to equation 33 from the \textsc{galfit} user's manual.\footnote{https://users.obs.carnegiescience.edu/peng/work/galfit/README.pdf} With a slight abuse of terminology, we define the gain $G$ as a scalar factor that converts any image units (in this case $e^{-} \, {\rm s}^{-1} \, {\rm pixel}^{-1}$) into $e^{-} \, {\rm pixel}^{-1}$.\footnote{The `true' gain is a factor that converts photon counts into electron counts, which is of order one for PS1 observations.} In the case of our mock POGS images, the gain $G$ is simply the total exposure time, for which we adopt typical values of 1200 and 750 s in the $i$ band and $g$ band, respectively.

\subsubsection{Creating a segmentation image}\label{subsubsec:segmap}

A segmentation image (also known as \textit{segmentation map}) is a 2D, integer-valued array with the same shape as the original image, where pixels belonging to the same source or `segment' are labelled with the same integer value. A value of zero is reserved for the background. Segmentation maps are often created with the popular software \textsc{SExtractor} \citep{Bertin1996}, which includes deblending of extended, overlapping sources, as well as a neural network algorithm to separate stars from galaxies. However, since our synthetic images contain a single galaxy each, simpler tools will suffice, as described next.

For each synthetic image, we create a segmentation image using the \texttt{photutils} photometry package.\footnote{https://photutils.readthedocs.io} We estimate the sky background by measuring the median of all pixel values, clipped iteratively at 3$\sigma$ until convergence, and then assuming that all pixel values 1.5$\sigma$ above the sky median belong to some source. We only keep the largest source since by construction there should be a single galaxy in each image. Finally, we smooth the main source segment by convolving the segmentation image with a uniform `boxcar' filter measuring 10 pixels in each dimension.

\section{Calculating optical morphologies}\label{sec:optical_morphologies}

To facilitate the analysis of our synthetic and survey images, we have developed \texttt{statmorph}, a Python package for calculating non-parametric morphologies of galaxy images, which we make publicly available.\footnote{https://statmorph.readthedocs.io} This tool is largely based on the IDL code described in \cite{Lotz2004, Lotz2006, Lotz2008, Lotz2008a} for calculating the Gini--$M_{20}$ \citep{Lotz2004} and concentration--asymmetry--smoothness \citep[CAS,][]{Conselice2003} statistics, with some additional features such as calculating multimode--intensity--deviation statistics \citep[MID,][]{Freeman2013} and fitting 2D S\'{e}rsic profiles. The code can handle images with a single source each, which is the mode used in this work, as well as large `mosaic' images with hundreds or thousands of sources. We have run \texttt{statmorph} on all of our Pan-STARRS galaxy images (Section \ref{subsec:obs_sample}), as well as on the synthetic images described in Section \ref{sec:synthetic_images}. The results are presented in Section \ref{sec:results}.

\begin{figure*}
  \centering
	\includegraphics[width=17.5cm]{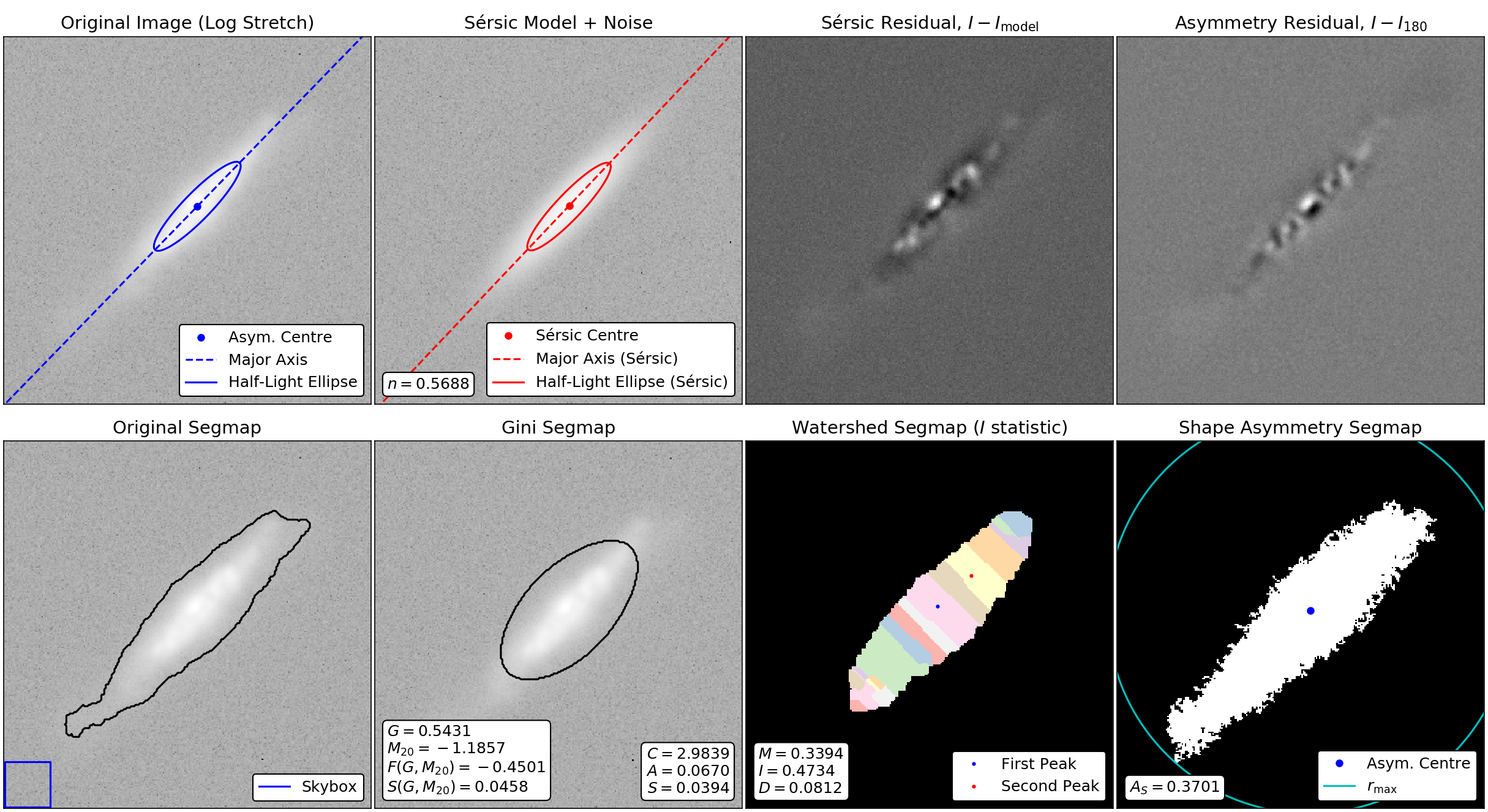}
	\caption{Some of the morphological measurements performed by \texttt{statmorph}, shown for the galaxy in the top row of Fig. \ref{fig:components} (using the full dust model). Top left: a cut-out from the original galaxy image, with pixel intensities scaled as $f(x) = \log(ax+1) / \log(a+1)$, where $a=10^4$. The blue point represents the point that minimizes the asymmetry (which is used as the galactic centre for most purposes), while the blue lines indicate the orientation and extent of the elliptical aperture that contributes half of the total flux. Top row, second from left: the fitted S\'{e}rsic model, with background noise added for realism. The S\'{e}rsic index, $n$, is indicated in the lower left-hand corner. The red point and lines show basic shape and size measurements derived from the S\'{e}rsic fit. Top row, third from left: the residual image obtained from subtracting the S\'{e}rsic model from the original image. Top right: the residual image obtained from subtracting the rotated image (with respect to the asymmetry centre) from the original image. Bottom left: the original image, with a black contour indicating the `galaxy' pixels identified by the original segmentation map. The blue square shows the `skybox' used to measure the properties of the background. Bottom row, second from left: the original image, with a black contour showing the pixels included in the calculation of the Gini--$M_{20}$ quantities. The text labels show the values of the Gini--$M_{20}$ and CAS statistics. Bottom row, third from left: the watershed segmentation image described in Section \ref{subsubsec:intensity}. Each coloured region indicates the pixels associated with each local maximum of the brightness distribution (by following the pixels' maximum gradient paths). The blue and red points indicate the brightest and second brightest maxima of the image. The text label shows the values of the MID statistics. Bottom right: the detection mask used to calculate the shape asymmetry, $A_S$. The cyan line shows the `minimal' circle that encloses the detection mask, which is used to define $r_{\rm max}$ (Section \ref{subsubsec:shape_asymmetry}).}
	\label{fig:morphology_measurements}
\end{figure*}

Fig. \ref{fig:morphology_measurements} shows some of the measurements carried out by \texttt{statmorph} for the $i$-band component of the simulated galaxy shown in the top panel of Fig. \ref{fig:components}, including full dust modelling. However, before describing in detail the measurements shown in Fig. \ref{fig:morphology_measurements} (Sections \ref{subsec:shape_measurements}--\ref{subsec:sersic_profiles}), we give a brief overview of the code next.

The \texttt{statmorph} code requires the following data:

\begin{itemize}

\item[(i)] $\tt image$: An image containing the object(s) of interest, given as a 2D array. \textit{The code assumes that the input image is already background subtracted.}

\item[(ii)] $\tt segmap$: The corresponding segmentation map, given as a 2D array (of the same size as the image) with different integer values used to label different sources. A value of zero is reserved for the background.

\item[(iii)] $\tt weightmap$ or $\tt gain$: The weight map is a 2D array (of the same size as the image) representing the 1$\sigma$ variation of each pixel value, including the contribution from the sky background.\footnote{We caution that, in general, different authors adopt different definitions of the weight map, which is sometimes expressed as the variance or the inverse variance (instead of the standard deviation) of the pixel values. Furthermore, it may or may not include the contribution from the background noise, and in some cases it may even refer to an exposure map (given in seconds).} If the weight map is not provided by the user, a `gain' parameter must be provided instead, i.e. a multiplicative factor that converts the image units into electron counts per pixel. This is used internally by the code to estimate the weight map assuming Poisson statistics.

\end{itemize}

The following two input arguments are optional, but their use is strongly recommended, if such data are available:

\begin{itemize}

\item[(iv)] $\tt mask$: A 2D array (of the same size as the image) with boolean values indicating the pixels that should be excluded from the calculation (e.g. pixels contaminated by foreground stars).

\item[(v)] $\tt psf$: A 2D array (usually smaller than the image) representing the PSF of the observations. This is only used for the S\'{e}rsic profile fitting.

\end{itemize}

Given the above information, \texttt{statmorph} will perform the morphological measurements described in Sections \ref{subsec:shape_measurements}--\ref{subsec:sersic_profiles}. Before doing so, however, we remove any `bad pixels' (as caused by cosmic rays or instrumental noise) from the image \citep{Lotz2004}, which we identify as pixels with flux values that deviate by more than $10 \sigma$ (this value can be modified by the user) from their neighbouring pixels.

In general, all of the `fixed' parameters implicit in the morphological measurements defined in Sections \ref{subsec:shape_measurements}--\ref{subsec:sersic_profiles} (e.g. the radius of the circular aperture used to measure the CAS statistics) can be modified by the user. For simplicity, however, we recommend leaving most parameters to their default values, which correspond to the ones most widely used in the literature.

After the calculations are done, \texttt{statmorph} will notify the user if there was a problem by means of the following two `bad measurement' flags, where values of `0' and `1' indicate good and bad measurements, respectively:

\begin{itemize}

\item $\tt flag$: This indicates if there was a problem with the basic measurements, which can happen for a variety of reasons (e.g. when there is some artifact, foreground star, or secondary source in the image that was not properly masked, or when the image is not background subtracted). Although the code will attempt to perform all of the morphological measurements regardless, we recommend only trusting the morphologies of galaxies that have $\tt flag == 0$.

\item $\tt flag\_sersic$: This indicates if there was a problem during the S\'{e}rsic profile fitting, which can happen for irregular or merging galaxies. Generally, it is not necessary to enforce this flag, unless one is actually interested in S\'{e}rsic profile fits.

\end{itemize}

\subsection{Shape measurements}\label{subsec:shape_measurements}

For every source labelled in the segmentation map, some basic shape measurements are performed. First, the centroid or barycentre $(x_{\rm c}, y_{\rm c})$ is calculated in the usual way:
\begin{flalign}
x_{\rm c} = \frac{\sum_{i,j}I_{ij} x_j}{\sum_{i,j} I_{ij}}, \hspace{0.5cm}
y_{\rm c} = \frac{\sum_{i,j}I_{ij} y_i}{\sum_{i,j} I_{ij}}, &&
\label{eq:centroid}
\end{flalign}
where $I_{ij}$ is the pixel value at the $i$th row and $j$th column of the image ${\bf I}$, $(x_j, y_i)$ are the coordinates at the centre of the pixel, and the sum is carried out over all the pixels belonging to the source, as specified by the input segmentation map.

In practice, the centroid does not always correspond to the `visual' centre of a galaxy, especially for merging systems or objects with extended tidal features. Therefore, the centroid is mostly used as an initial guess for more robust measures of the galactic centre, such as the point that minimizes the asymmetry index $A$, which is described in Section \ref{subsubsec:asymmetry}.

For either choice of the galactic centre, all central moments up to second order are calculated:
\begin{flalign}
\mu_{pq} = \sum_{i,j} I_{ij} (x_j - x_{\rm c})^p (y_i - y_{\rm c})^q, &&
\label{eq:central_moments}
\end{flalign}
where $p, q \in \{0, 1, 2\}$. In particular, the second-order moments are used to construct the covariance matrix of the light distribution ${\bf I}$:
\begin{flalign}
{\rm cov}({\bf I}) = \frac{1}{\mu_{00}} \left(\begin{matrix*}[c]
\mu_{20} & \mu_{11} \\
\mu_{11} & \mu_{02}
\end{matrix*}\right) \equiv \left(\begin{matrix*}[c]
\overline{x^2} & \overline{xy} \\
\overline{xy} & \overline{y^2}
\end{matrix*}\right). &&
\label{eq:covariance_matrix}
\end{flalign}
This completely determines a 2D Gaussian distribution with second moments equal to those of the original source. The orientation of the source, i.e. the angle between the $x$-axis and the major axis of the corresponding Gaussian distribution, is given by
\begin{flalign}
{\rm orientation} = \frac{1}{2} \arctan \left(\frac{2 \, \overline{xy}}{\overline{x^2} - \overline{y^2}}\right). &&
\label{eq:orientation}
\end{flalign}

Similarly, if $\lambda_1$ and $\lambda_2$ are the eigenvalues of the covariance matrix (\ref{eq:covariance_matrix}), where $\lambda_1 \geq \lambda_2$, then the elongation and ellipticity of the source can be calculated as
\begin{flalign}
{\rm elongation} = \frac{a}{b}&&
\label{eq:elongation}
\end{flalign}
and
\begin{flalign}
{\rm ellipticity} = 1 - \frac{b}{a},&&
\label{eq:ellipticity}
\end{flalign}
where $a \equiv \sqrt{\lambda_1}$ and $b \equiv \sqrt{\lambda_2}$. Equivalently, closed-form solutions for $a$ and $b$ can be found in equations 24--25 from the \textsc{SExtractor} user's manual.\footnote{https://www.astromatic.net/software/sextractor}

Finally, we note that an alternative shape measurement included in \texttt{statmorph} is the ellipticity derived from S\'{e}rsic profile fitting, which is described in Section \ref{subsec:sersic_profiles}.

\subsection{Size measurements}\label{subsec:size_measurements}

The main size measurements carried out by \texttt{statmorph} are the half-light radius $r_{\rm half}$ and the Petrosian radius $r_{\rm petro}$ \citep{Petrosian1976}, although other size measurements are necessary for some of the morphological statistics described next, and are therefore included as well. This includes $r_{20}$ and $r_{80}$ (Section \ref{subsubsec:concentration}), $r_{\rm max}$ (Section \ref{subsubsec:shape_asymmetry}), and the S\'{e}rsic half-light radius (Section \ref{subsec:sersic_profiles}).

Most of our size measurements are carried out for both circular and elliptical apertures (using the orientation and ellipticity from Section \ref{subsec:shape_measurements}). For the sake of readability, we will often use the term `radius' both for the radius of a circle and for the semi-major axis of an ellipse, unless noted otherwise.

\subsubsection{Half-light radius}\label{subsubsec:rhalf}

The half-light radius, $r_{\rm half}$, contains half of the light emitted by the galaxy. We calculate it both for circular and elliptical apertures, setting the centre at the point that minimizes the asymmetry index $A$ (Section \ref{subsubsec:asymmetry}). There is some ambiguity in defining the outer radius, which is assumed to contain the galaxy's total flux. In an attempt to capture as much of the galaxy's light as possible, we define the outer radius at $r_{\rm max}$ (Section \ref{subsubsec:shape_asymmetry}). However, we note that this definition of $r_{\rm half}$ is sensitive to the limiting surface brightness and S/N of the observations. As an alternative, we describe $r_{50}$ in Section \ref{subsubsec:concentration}, which represents a measure of the half-light radius consistent with $r_{20}$ and $r_{80}$ from the CAS statistics. Finally, in Section \ref{subsec:sersic_profiles} we present the S\'{e}rsic half-light radius, which contains half of the light of a fitted 2D S\'{e}rsic profile.

\subsubsection{Petrosian radius}\label{subsubsec:rpetro}

The Petrosian radius \citep{Petrosian1976}, $r_{\rm petro}$, is the radius at which the mean surface brightness is equal to some fraction $\eta$ of the mean surface brightness within $r_{\rm petro}$. The Petrosian radius is largely insensitive to variations in the limiting surface brightness and S/N of the observations, which makes it ideal for comparing galaxy properties across different redshifts. Typically, the Petrosian `ratio' is set to $\eta = 0.2$. We calculate the Petrosian radius both for circular and elliptical apertures, assuming that the centre coincides with the point that minimizes the asymmetry index $A$ (Section \ref{subsubsec:asymmetry}).

\subsection{Noise and flux measurements}\label{subsec:noise_and_flux_measurements}

\subsubsection{Sky background properties}\label{subsubsec:skybox}

It is generally useful to quantify some properties of the sky background, which is defined as the set of pixels that do not belong to any source (i.e. pixels with a value of zero in the segmentation map) and that are also not masked.  For most purposes, it is sufficient to simply search for a square region in the image that contains only background pixels \citep[however, see][for a more sophisticated approach]{Shi2009}. Our code finds such a `skybox' automatically, searching first for regions measuring $32 \times 32$ pixels and iteratively halving the dimensions of the skybox candidates until a blank region of the image is found. Once an appropriate skybox has been determined, the mean, median and standard deviation of its pixel flux values are calculated. The skybox is also used in Section \ref{subsec:cas_statistics} to measure the asymmetry and smoothness of the background.

\subsubsection{S/N per pixel}\label{subsubsec:sn_per_pixel}

As mentioned at the beginning of Section \ref{sec:optical_morphologies}, the weight map quantifies the amount of variation in each pixel value. Therefore, we calculate the average S/N per pixel, $\left\langle {\rm S/N} \right\rangle$, as the average ratio between the image and the weight map over some aperture. For consistency with \cite{Lotz2004}, such an aperture is given by the Gini segmentation map defined in Section \ref{subsubsec:gini_coefficient}, which roughly corresponds to an elliptical aperture at the Petrosian radius.

\subsubsection{Petrosian flux}\label{subsubsec:petrosian_flux}

In order to estimate galaxy magnitudes and colours, it is useful to output the sum of the pixel flux values over some aperture. Here, we follow the SDSS convention and measure the Petrosian flux over an aperture with a radius equal to \textit{twice} the Petrosian radius \citep{Stoughton2002}. We measure the Petrosian flux both for circular and elliptical apertures.

\subsection{Gini--$M_{20}$ statistics}

The Gini--$M_{20}$ classification system \citep{Lotz2004} has been used extensively to quantify galaxy morphologies, including not just `normal' galaxies but also irregular and merging galaxies. In particular, \cite{Lotz2008a} defined a simple partitioning of the Gini--$M_{20}$ diagram that can be used to roughly separate early-type, late-type, and merging galaxies at $0.2 < z < 0.4$. A generalization of this partitioning scheme leads to the definition of the Gini--$M_{20}$ `bulge statistic' (Section \ref{subsubsec:bulge_statistic}) and the Gini--$M_{20}$ `merger statistic' (Section \ref{subsubsec:merger_statistic}).

\subsubsection{Gini coefficient}\label{subsubsec:gini_coefficient}

The Gini coefficient ($G$) is a statistic traditionally used in economics to quantify the wealth inequality in a population, which has recently found an application in astronomy as well \citep{Abraham2003, Lotz2004}. For a set of $n$ pixel flux values $X_i$, where $i = 1, 2, ..., n$, the Gini coefficient can be computed as \citep{Glasser1962}
\begin{flalign}
G = \frac{1}{\bar{X} n (n-1)} \sum_{i=1}^{n} (2i - n - 1) X_i,&&
\label{eq:gini1}
\end{flalign}
where $\bar{X}$ represents the mean over the pixel values. A value of $G=1$ is obtained when all of the flux is concentrated in a single pixel, while a homogeneous brightness distribution yields $G=0$.

In practice, the Gini coefficient is sensitive to the choice of pixels that are included in the calculation. Originally, \cite{Abraham2003} considered pixels that lie above a constant surface brightness threshold, which makes a comparison between low- and high-redshift galaxies difficult. Therefore, \cite{Lotz2004} proposed a method to construct a `Gini' segmentation map that depends only on the Petrosian radius, which is therefore insensitive to the $(1+z)^4$ surface brightness dimming of distant galaxies. This segmentation map is created in the following way. First, the galaxy image is convolved with a Gaussian kernel with $\sigma = r_{\rm petro} / 5$, where $r_{\rm petro}$ is the elliptical Petrosian `radius' (i.e. the semi-major axis). Then, the mean surface brightness at $r_{\rm petro}$ is used to define a flux threshold, so that pixels with flux values above this threshold are assigned to the galaxy. Note that while the shape of the Gini segmentation map is quite smooth by construction, $G$ is always calculated for the original, unsmoothed image.\footnote{We also point out that the Gini segmentation map is generally expected to be continuous: when this is not the case, the `bad measurement' flag is activated, although $G$ will be calculated anyway for the main individual segment.}

\cite{Lotz2004} also noted that the traditional definition of the Gini coefficient, as given by eq. (\ref{eq:gini1}), becomes unreliable for very noisy galaxy images, especially at $\left\langle {\rm S/N} \right\rangle \lesssim 3$, due to the inclusion of increasingly negative pixel flux values. They found that a first-order correction consists in calculating $G$ for the absolute values of the pixel flux values, effectively redefining the Gini coefficient as
\begin{flalign}
G = \frac{1}{\overline{|X|} n (n-1)} \sum_{i=1}^{n} (2i - n - 1) |X_i|,&&
\label{eq:gini2}
\end{flalign}
where $\overline{|X|}$ is the mean of the absolute values $|X_i|$. This formula can recover the `true' Gini coefficient to within 10 per cent for galaxies with $\left\langle {\rm S/N} \right\rangle = 2$--$3$ \citep{Lotz2004}.

\subsubsection{$M_{20}$}\label{subsubsec:m20}

The $M_{20}$ statistic \citep{Lotz2004} measures the second moment of a galaxy's brightest regions, containing 20 per cent of the total flux, relative to the total second-order central moment, $\mu_{\rm tot}$. The latter is calculated as
\begin{flalign}
\mu_{\rm tot} = \sum_{i=1}^n \mu_i \equiv \sum_{i=1}^n I_i \left[(x_i - x_{\rm c})^2 + (y_i - y_{\rm c})^2\right],  &&
\label{eq:mtot}
\end{flalign}
where $I_i$ are the pixel flux values, $(x_{\rm c}, y_{\rm c})$ is the galaxy's centre, and the sum is carried out over all the pixels identified by the Gini segmentation map (Section \ref{subsubsec:gini_coefficient}). The centre $(x_{\rm c}, y_{\rm c})$ is the point that minimizes $\mu_{\rm tot}$, which corresponds to the centroid of the pixels labelled in the segmentation map.

The second moment of the galaxy's brightest regions is obtained by sorting the pixels by flux and summing $\mu_i$ over the brightest pixels until the sum of the brightest pixels equals 20 per cent of the galaxy's total flux. The result is then normalized by $\mu_{\rm tot}$, so that $M_{20}$ is ultimately defined as \citep{Lotz2004}:
\begin{flalign}
M_{20} \equiv \log_{10}\left(\frac{\sum_i \mu_i}{\mu_{\rm tot}}\right), \;\;
{\rm while} \;\;
\sum_i I_i \, < \, 0.2 I_{\rm tot}, &&
\label{eq:m20}
\end{flalign}
where $I_{\rm tot}$ is the total flux of the pixels identified by the segmentation map.

\subsubsection{The bulge statistic}\label{subsubsec:bulge_statistic}

\cite{Lotz2008a} classified $0.2 < z < 0.4$ merging galaxies as objects on the $G$--$M_{20}$ diagram such that $G > -0.14M_{20} + 0.33$, and defined early-type galaxies (including E/S0/Sa Hubble types) as non-merging galaxies that also satisfy $G > 0.14M_{20} + 0.80$. Note that the corresponding lines on the $G$--$M_{20}$ diagram have slopes of $-0.14$ and $0.14$, respectively, and that they intersect at
\begin{flalign}
G_0 = 0.565, \hspace{0.5cm} M_{20,0} = -1.679.&&
\label{eq:gini_m20_origin}
\end{flalign}

The Gini--$M_{20}$ `bulge statistic' \citep{Snyder2015} is defined as the position along a line (with values increasing towards bulge-dominated systems) with origin\footnote{The original definition by \cite{Snyder2015} placed the origin at a slightly different position: $G_0 = 0.533$, $M_{20,0} = -1.75$. However, since we are only interested in relative values, and because $F(G, M_{20})$ is a linear combination of $G$ and $M_{20}$, the location of the origin is irrelevant for our purposes.} at $(G_0, M_{20,0})$ that is perpendicular to the line separating early-type and late-type galaxies, scaled by a factor of 5:
\begin{flalign}
F(G, M_{20}) = -0.693 M_{20} + 4.95 G - 3.96. &&
\label{eq:bulge_statistic}
\end{flalign}

This quantity is strongly correlated with other measures of galactic bulge strength, such as the concentration index (Section \ref{subsubsec:concentration}) and the S\'{e}rsic index (Section \ref{subsec:sersic_profiles}).

\subsubsection{The merger statistic}\label{subsubsec:merger_statistic}

Similarly, the Gini--$M_{20}$ `merger statistic' \citep{Snyder2015a} is the position along a line (with values increasing towards merging systems) with origin at $(G_0, M_{20,0})$ that is perpendicular to the line that divides merging and non-merging galaxies:
\begin{flalign}
S(G, M_{20}) = 0.139 M_{20} + 0.990 G - 0.327. &&
\label{eq:merger_statistic}
\end{flalign}

We note that \cite{Lotz2004} originally defined slightly different division lines, tuned to observations of $z=0$ galaxies (instead of the more distant $0.2 < z < 0.4$ galaxies considered in \citealt{Lotz2008a}), which would perhaps be a better choice for this work (we consider Pan-STARRS galaxies at $z \sim 0.05$). Instead, we adopt the classification scheme of \cite{Lotz2008a} in order to provide a consistent, `general-purpose' definition of the $G$-$M_{20}$ bulge and merger statistics that can be applied over a wider range of redshifts \citep{Snyder2015a, Snyder2015}.

\subsection{\textit{CAS} statistics}\label{subsec:cas_statistics}

Another set of widely used non-parametric morphological indicators consists of the CAS statistics. Although galaxy concentration has been measured as early as \cite{Morgan1958, Morgan1959}, while the asymmetry and smoothness indices have predecessors in the works of \cite{Rix1995} and \cite{Takamiya1999}, respectively, these quantities have evolved over the years, until reaching their current form as described in \cite{Conselice2003}.

\subsubsection{Concentration}\label{subsubsec:concentration}

The concentration index ($C$) is usually defined as \citep{Bershady2000, Conselice2003}:
\begin{flalign}
C = 5\log_{10}\left(\frac{r_{80}}{r_{20}}\right), &&
\label{eq:concentration}
\end{flalign}
where $r_{20}$ and $r_{80}$ are the radii of circular apertures containing 20 and 80 per cent, respectively, of the galaxy's light.\footnote{For convenience, we also calculate $r_{50}$ in an analogous fashion, which can be considered as an alternative to the definition of $r_{\rm half}$ presented in Section \ref{subsubsec:rhalf}.} There is some ambiguity in defining the aperture that contains the `total' flux of a galaxy. \cite{Bershady2000} defined the total flux as the flux contained within an aperture with radius equal to $2 r_{\rm petro}$. However, more recent studies \citep{Conselice2003, Lotz2004} measure the total flux within $1.5 r_{\rm petro}$, which is also the definition that we adopt here. The centre of the aperture corresponds to the point that minimizes the asymmetry index $A$, described next.

\subsubsection{Asymmetry}\label{subsubsec:asymmetry}

The asymmetry index ($A$) is obtained by subtracting the galaxy image rotated by $180^{\circ}$ from the original image \citep{Schade1995, Abraham1996, Conselice2000}:
\begin{flalign}
A = \frac{\sum_{i,j}|I_{ij} - I_{ij}^{180}|}{\sum_{i,j}|I_{ij}|} - A_{\rm bgr}, &&
\label{eq:asymmetry}
\end{flalign}
where $I_{ij}$ and $I_{ij}^{180}$ are the pixel flux values of the original and rotated images, respectively, and $A_{\rm bgr}$ is the average asymmetry of the background. The sum is carried out over all pixels within $1.5 r_{\rm petro}$ of the galaxy's centre, which is determined by minimizing $A$.

\subsubsection{Smoothness}\label{subsubsec:smoothness}

The `clumpiness' or smoothness index ($S$) is obtained by subtracting the galaxy image smoothed with a boxcar filter of width $\sigma$ from the original image \citep{Conselice2003}:
\begin{flalign}
S = \frac{\sum_{i,j}I_{ij} - I_{ij}^{S}}{\sum_{i,j}I_{ij}} - S_{\rm bgr}, &&
\label{eq:smoothness}
\end{flalign}
where $I_{ij}$ and $I_{ij}^{S}$ are the pixel flux values of the original and smoothed images, respectively, and $S_{\rm bgr}$ is the average smoothness of the background. The sum is carried out over all pixels at distances between $\sigma$ and $1.5 r_{\rm petro}$ from the point that minimizes the asymmetry index $A$. The central region is excluded because most galaxies are highly concentrated. In addition, pixels such that $I_{ij} - I_{ij}^{S} < 0$ are excluded from the numerator in eq. (\ref{eq:smoothness}), so that only high-frequency features that are \textit{brighter} than the local average contribute to $S$. Following \cite{Lotz2004}, we set $\sigma = 0.25 r_{\rm petro}$ (instead of the original choice of $\sigma = 0.3 r_{\rm petro}$). Note that larger values of $S$ actually correspond to galaxies that are \textit{less} smooth (i.e. more `clumpy').

\subsection{\textit{MID} statistics}\label{subsec:mid_statistics}

The MID statistics \citep[][]{Freeman2013, Peth2016} were introduced as an alternative to the Gini--$M_{20}$ and CAS statistics that is plausibly more sensitive to recent mergers. However, these new statistics have not been tested as extensively as the Gini--$M_{20}$ and CAS statistics, especially using hydrodynamic simulations \citep[e.g.][]{Lotz2008, Lotz2010, Lotz2010a, Bignone2017}.

The MID statistics are calculated over a segmentation map that can be considered to be a generalization of the `Gini' segmentation map defined in Section \ref{subsubsec:gini_coefficient}, removing the assumption of ellipticity. It is constructed by finding a surface brightness threshold such that its value equals a fraction $\eta$ of the mean surface brightness of the pixels above the threshold, where typically $\eta = 0.2$. The main source is then identified as the set of connected pixels (including the brightest pixel) with flux values above the threshold. The shape of the segmentation map is then regularized slightly with a $3 \times 3$ boxcar filter.

\subsubsection{Multimode}\label{subsubsec:multimode}

The multimode statistic ($M$) measures the ratio between the areas of the two most `prominent' clumps within a galaxy. Its calculation mostly consists in finding such substructures, which is done as follows. First, all pixels within the MID segmentation map are sorted by brightness. Then, for a given quantile $q$ (between 0 and 1), the set of all pixels with flux values above the $q$th quantile will generally consist of $n$ groups of contiguous pixels, which are sorted by area (largest first): $A_{q,1}$, $A_{q,2}$, ..., $A_{q,n}$. Finally, $M$ is defined as the quantile $q$ that maximizes the area ratio\footnote{In the original definition by \cite{Freeman2013}, the area `ratio' included an additional factor of $A_{q,2}$, which introduces a dependence on the size of the galaxy. This was changed by \cite{Peth2016}, which is the definition followed in this work.} between the two largest groups \citep{Peth2016}:
\begin{flalign}
M = \max_{q}\left(\frac{A_{q,2}}{A_{q,1}}\right). &&
\label{eq:multimode}
\end{flalign}

\subsubsection{Intensity}\label{subsubsec:intensity}

The intensity statistic ($I$) measures the ratio between the two brightest subregions of a galaxy. To calculate it, the galaxy image is first slightly smoothed using a Gaussian kernel with $\sigma = 1$ pixel. Then, the image is partitioned into pixel groups according to the watershed algorithm: each distinct subregion consists of all the pixels such that their maximum gradient paths lead to the same local maximum. We perform this operation using a function from the \texttt{scikit-image} image-processing package.\footnote{https://scikit-image.org} A watershed segmentation image is illustrated in Fig. \ref{fig:morphology_measurements}. Once the pixel groups are defined, their summed intensities are sorted into descending order: $I_1$, $I_2$, etc. The intensity statistic is then defined as \citep{Freeman2013}
\begin{flalign}
I = \frac{I_2}{I_1}. &&
\label{eq:intensity}
\end{flalign}

\subsubsection{Deviation}\label{subsubsec:deviation}

The deviation statistic ($D$) measures the distance between the image centroid, $(x_{\rm c}, y_{\rm c})$, calculated for the pixels identified by the MID segmentation map, and the brightest peak found during the computation of the $I$ statistic, $(x_{I_1}, y_{I_1})$. This distance is normalized by $\sqrt{n_{\rm seg}/\uppi}$, where $n_{\rm seg}$ is the number of pixels in the segmentation map, which represents an approximate galaxy `radius' \citep{Freeman2013}:
\begin{flalign}
D = \sqrt{\frac{\uppi}{n_{\rm seg}}} \sqrt{(x_{\rm c} - x_{I_1})^2 + (y_{\rm c} - y_{I_1})^2}. &&
\label{eq:deviation}
\end{flalign}

\subsection{Other non-parametric morphologies}\label{subsec:other_nonparametric}

\subsubsection{Outer asymmetry}\label{subsubsec:outer_asymmetry}

The outer asymmetry ($A_{\rm O}$) is similar to the standard asymmetry, $A$ (Section \ref{subsubsec:asymmetry}), but its calculation excludes pixels within the inner elliptical aperture that contributes 50 per cent of the galaxy's light \citep{Wen2014}. More precisely, the flux integration is carried out over an elliptical annulus with inner and outer semi-major axes $a_{\rm half}$ and $a_{\rm max}$, which are the elliptical generalizations of $r_{\rm half}$ (Section \ref{subsubsec:rhalf}) and $r_{\rm max}$ (Section \ref{subsubsec:shape_asymmetry}), respectively. The orientation and ellipticity of the annulus are calculated as described in Section \ref{subsec:shape_measurements}. \cite{Wen2014} originally rotated each image around the centroid of the outer elliptical annulus. However, for similar reasons to those explained at the end of Section \ref{subsubsec:shape_asymmetry}, we instead rotate each galaxy image around the point that minimizes the standard asymmetry, $A$.

\subsubsection{Shape asymmetry}\label{subsubsec:shape_asymmetry}

The shape asymmetry ($A_{\rm S}$) is calculated using the same mathematical expression as the standard asymmetry (eq. \ref{eq:asymmetry}). However, the measurement is done on the binary detection mask (i.e. the segmentation map) rather than the galaxy image. This increases the sensitivity to low surface brightness features in the galaxy's outskirts \citep{Pawlik2016}. The calculation of $A_{\rm S}$ mostly consists in carefully separating the `galaxy' pixels from the background.

The binary detection mask or `shape asymmetry segmentation map' is created as follows. First, the sky background level is estimated over a circular annulus with inner and outer radii equal to 2 and 4 times the Petrosian semi-major axis (Section \ref{subsubsec:rpetro}).\footnote{Instead, \cite{Pawlik2016} used an annulus with inner and outer radii equal to 20 and 40 times the FWHM of the galaxy profile.} Within this aperture, the sky background level is estimated by iteratively clipping the histogram of the flux pixel values at $3 \sigma$ until convergence of the `mode' defined in \cite{Bertin1996}: ${\rm mode = 2.5 \times median - 1.5 \times mean}$. Then, a brightness threshold is defined at $1 \sigma$ above the mode and the galaxy image is smoothed slightly with a $3 \times 3$ boxcar (mean) filter. The binary detection mask is defined as the contiguous group of pixels above the threshold that includes the brightest pixel in the galaxy image. An example is shown in the bottom right-hand panel of Fig. \ref{fig:morphology_measurements}.

Once the detection mask has been created, we define $r_{\rm max}$ as the distance between the point that minimizes the standard asymmetry ($A$) and the most distant pixel that belongs to the detection mask.\footnote{We differ slightly from \cite{Pawlik2016}, who measured $r_{\rm max}$ with respect to the brightest pixel instead of the asymmetry centre.} This defines a `minimal' circular aperture that encloses the detection mask.\footnote{For completeness, we also calculate the elliptical generalization of $r_{\rm max}$, i.e. the semi-major axis of the minimal elliptical aperture (with shape defined as in Section \ref{subsec:shape_measurements}) that contains the binary mask.} As shown by \cite[][their fig. 2]{Pawlik2016}, even a circular aperture at $2 r_{\rm petro}$ does not capture the low surface brightness features of some morphologically disturbed galaxies, which makes $r_{\rm max}$ a more useful size measurement in this context.

Finally, $A_{\rm S}$ is calculated by applying eq. (\ref{eq:asymmetry}) to the binary detection mask (note that the background correction term disappears since it is zero by definition). However, instead of finding the point that minimizes $A_{\rm S}$, the rotation is done around the point that minimizes the standard (i.e. flux-weighted) asymmetry, $A$. This ensures that the galaxy is rotated around its core, rather than around some arbitrary point that could be far from the `visual' centre of the galaxy, especially for strongly disturbed systems.

\subsection{S\'{e}rsic profile fitting}\label{subsec:sersic_profiles}

Although \texttt{statmorph} was originally created to calculate non-parametric morphologies, we have decided to include S\'{e}rsic profile fits as well, mostly due to their importance in astronomy. The S\'{e}rsic profile \citep{Sersic1968} is a useful parametrization of the brightness profile of a galaxy:
\begin{flalign}
I(r) = I_{e} \exp\left\{ -b_n \left[ \left(\frac{r}{r_e}\right)^{1/n} - 1\right] \right\}, &&
\label{eq:sersic_profile}
\end{flalign}
where $n$ is known as the S\'{e}rsic index, $b_n$ is a coefficient chosen so that a circular aperture with radius $r_e$ contains half of the galaxy's flux, and $I_e$ is the surface brightness at $r=r_e$. Note that the exponential profile (which approximates the brightness profile of spiral galaxies) and the de Vaucouleurs profile (typical of elliptical galaxies) correspond to the special cases $n=1$ and $n=4$, respectively.

Although eq. (\ref{eq:sersic_profile}) defines a 1D profile (i.e. an azimuthally symmetric brightness profile), it can also be used to represent a 2D profile -- with elliptical isophotes -- through an appropriate transformation of the coordinate system. In total, a 2D S\'{e}rsic profile has 7 free parameters: $I_e$, $r_e$, $n$, the centre of the profile (which contributes two free parameters), and the ellipticity and orientation of the elliptical isophotes.

There are highly sophisticated algorithms available for fitting S\'{e}rsic profiles to astronomical images, such as \textsc{gim2d} \citep{Simard1998}, \textsc{galfit} \citep{Peng2002}, \textsc{imfit} \citep{Erwin2015}, and \textsc{profit} \citep{Robotham2017}, among others. These codes typically provide the option of fitting other mathematical models besides the S\'{e}rsic profile, and support fitting multiple components simultaneously. However, since we are only interested in fitting a single-component S\'{e}rsic profile to each galaxy image, simpler tools are sufficient for our purposes, as summarized next.

We fit a 2D S\'{e}rsic profile to each galaxy image using the \texttt{astropy}\footnote{http://www.astropy.org} modelling package, which in turn uses fitting functions from other open source projects, in particular \texttt{scipy}.\footnote{https://www.scipy.org} The fitting is done with the Levenberg--Marquardt algorithm, assigning a weight to each individual pixel given by the weight map described at the beginning of Section \ref{sec:optical_morphologies}. During each step of the fitting, the modelled S\'{e}rsic profile is convolved with the PSF of the observations, if provided. The initial `guessed' values for the fitted parameters are based on the non-parametric measurements presented in Sections \ref{subsec:shape_measurements} through \ref{subsec:other_nonparametric}, and are generally not far from the final, optimized values.

\section{Results}\label{sec:results}

\begin{figure*}
	\centerline{
	  \includegraphics[width=17.5cm]{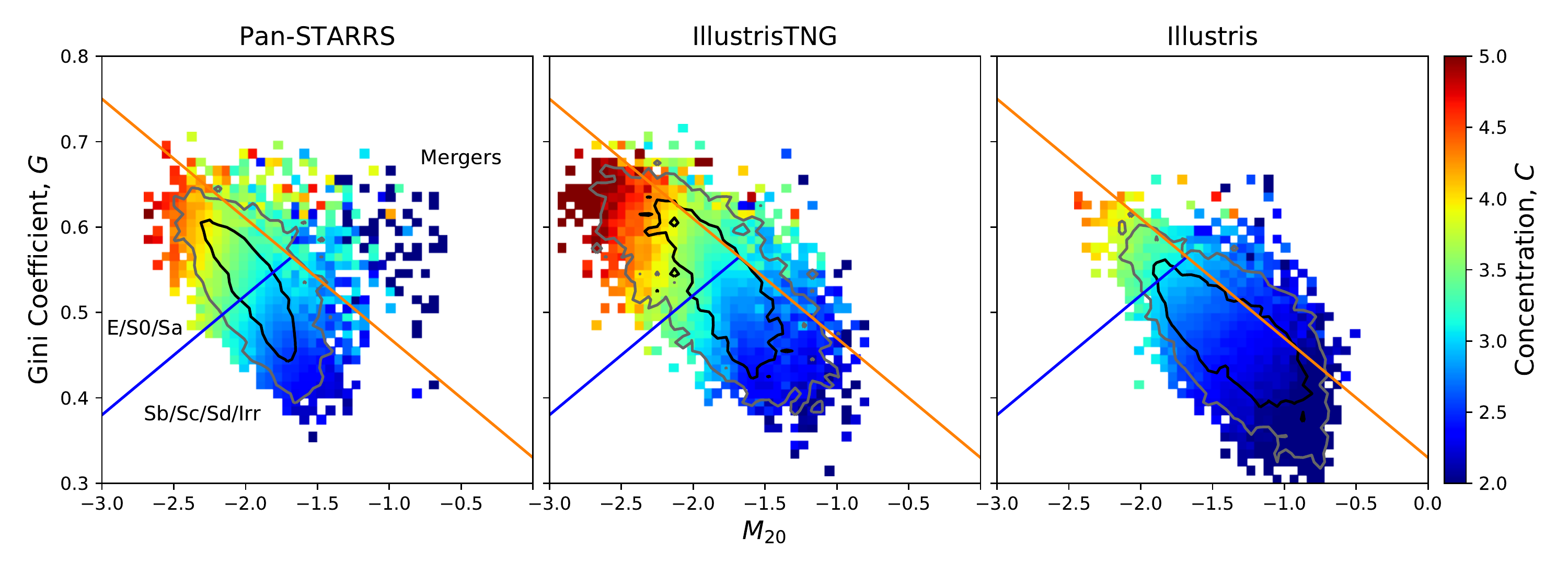}
	}
	\caption{The Gini--$M_{20}$ diagram for our galaxy sample, consisting of objects with stellar masses $\log_{10}(M_{\ast}/\Msun) \approx 9.8$--$11.3$ at $z \approx 0.05$. The left-hand panel shows galaxies observed with Pan-STARRS, while the middle and right-hand panels show galaxies from the IllustrisTNG and original Illustris simulations, respectively. The Gini--$M_{20}$ diagram is coloured according to the median value (in each 2D bin) of the concentration index, $C$. The black and dark grey contours contain 68 and 95 per cent of the galaxy distribution. The orange and blue lines were defined in \protect\cite{Lotz2008a} to roughly separate merger candidates (above the orange line) from early-type (above the blue line) and late-type (below the blue line) galaxies at $0.2 < z < 0.4$. These lines are also used to define the $G$-$M_{20}$ `bulge' and 'merger' statistics (Sections \ref{subsubsec:bulge_statistic} and \ref{subsubsec:merger_statistic}). Overall, the locus of the $G$-$M_{20}$ diagram in IllustrisTNG and the associated $C$ values are consistent with observations. The apparent lack of merging systems in IllustrisTNG is simply a consequence of the synthetic image generation procedure, which considers a single \textsc{subfind} object at a time, effectively biasing against early-stage mergers.}
	\label{fig:gini_m20}
\end{figure*}

Throughout this section, we present the results of performing the morphology measurements described in Section \ref{sec:optical_morphologies} on the synthetic images described in Section \ref{sec:synthetic_images}, as well as on our sample of real Pan-STARRS images (Section \ref{subsec:obs_sample}). We only show those measurements that satisfy the following properties:

\begin{itemize}
\item[(i)] They should be classified as `good' measurements, i.e. they should satisfy \texttt{flag == 0} (Section \ref{sec:optical_morphologies}). For parameters derived from S\'{e}rsic fitting we also require \texttt{flag\_sersic == 0}.
\item[(ii)] The mean S/N per pixel, $\left\langle {\rm S/N} \right\rangle$, should be higher than 2.5 \citep{Lotz2006}.
\item[(iii)] The radius of the circular aperture containing 20 per cent of the galaxy's flux, $r_{\rm 20}$, should be larger than half of the FWHM of the PSF.
\end{itemize}

We find that $\sim$90 per cent of our galaxy sample satisfies these requirements in any stellar mass range, except for real Pan-STARRS galaxies at $M_{\ast} \gtrsim 10^{11} \, \Msun$, where the percentage drops to $\sim$80 per cent. All of the morphological measurements presented throughout this section were obtained in the PS1 $i$ band.

We highlight that the mean FWHM of PS1 $i$-band observations is 1.11 arcsec, which corresponds to $\sim$1 kpc at $z \sim 0.05$. On the other hand, the spatial resolution of the Illustris and IllustrisTNG simulations, essentially set by the gravitational softening length, is slightly below $\sim$1 kpc, which is consistent with the observational sample. This also happens to be approximately the spatial resolution limit at which structural measurements such as the Gini coefficient ($G$), $M_{20}$, the concentration index ($C$), the asymmetry index ($A$), and the clumpiness index ($S$) are still considered to be reliable \citep{Lotz2006}.

\subsection{Overview}\label{subsec:overview}

\begin{figure*}
  \centering
	\includegraphics[width=17.5cm]{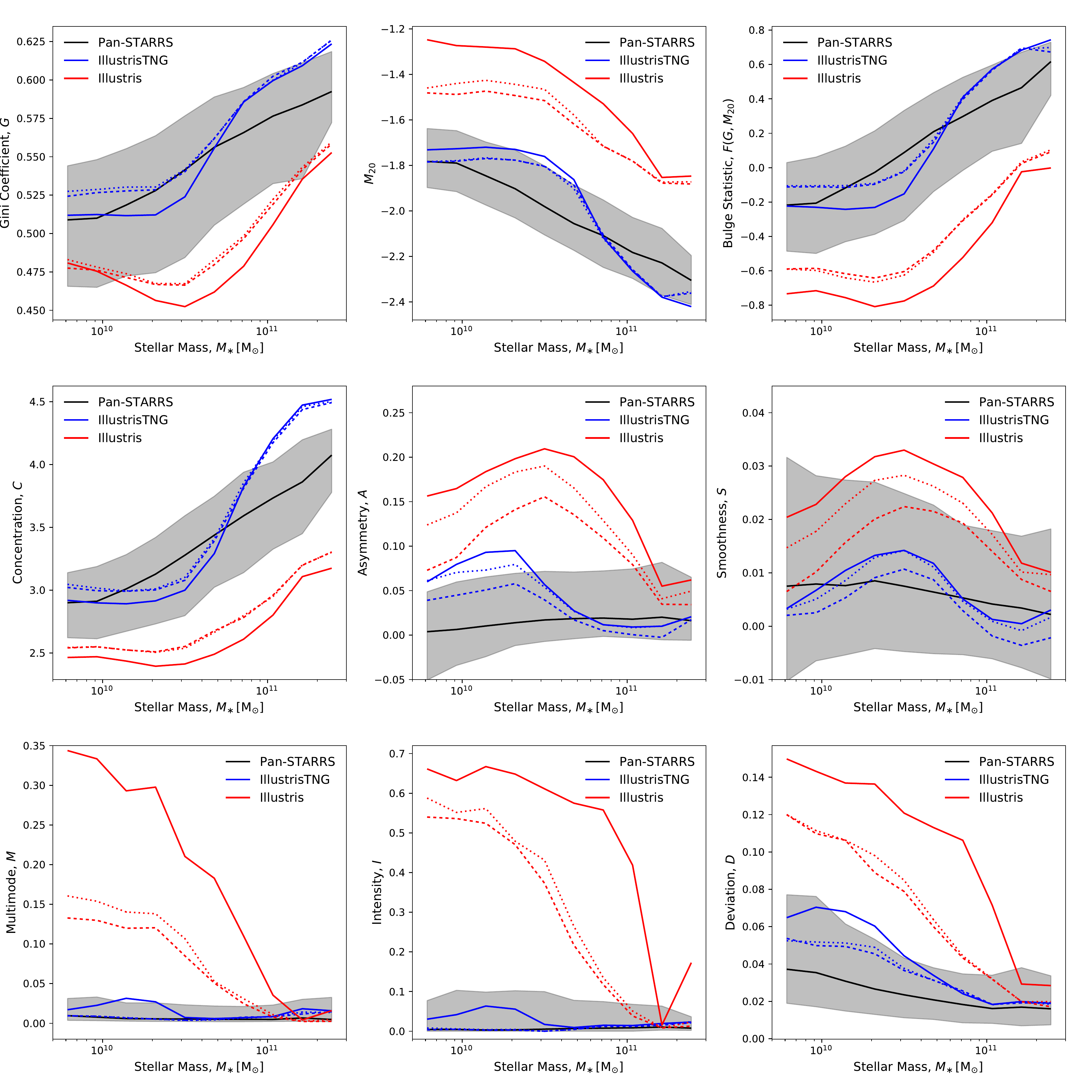}
	\caption{Median trends of various morphological parameters, as indicated in the $y$-axis labels, as a function of stellar mass. The top, middle and bottom rows show the Gini--$M_{20}$, CAS and MID statistics, respectively. In each panel, the black line shows the median trend for the observational Pan-STARRS sample, while the grey shaded region indicates the 16th to 84th percentile range of the same data. The blue and red lines show results from the IllustrisTNG and original Illustris simulations, respectively, with different line styles indicating model variations: the solid lines show results obtained with the \textsc{skirt} radiative transfer code including full dust modelling, while the dashed and dotted lines represent simpler models with an unresolved dust distribution and without dust, respectively. This figure shows that the optical morphologies of IllustrisTNG galaxies are in better overall agreement with observations than those of the original Illustris simulation.}
	\label{fig:morph_vs_mstar_1}
\end{figure*}

\begin{figure*}
  \centering
	\includegraphics[width=17.5cm]{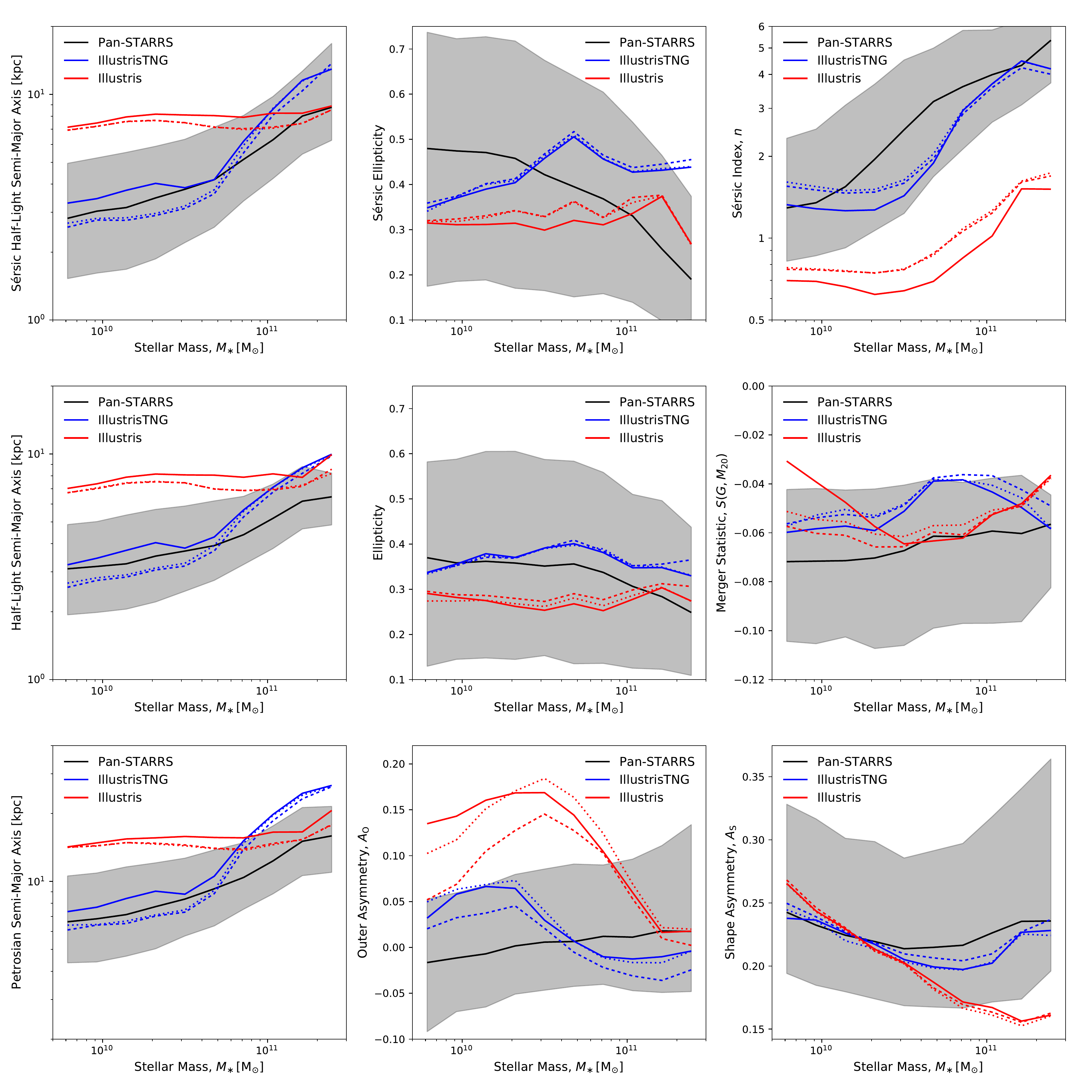}
	\caption{Same as Fig. \ref{fig:morph_vs_mstar_1}, but this time showing S\'{e}rsic parameters (top row), size measurements (left-hand column), shape measurements (top centre and middle panels), the Gini--$M_{20}$ merger statistic (centre right-hand panel), and alternative definitions of the asymmetry index (bottom centre and bottom right-hand panels).}
	\label{fig:morph_vs_mstar_2}
\end{figure*}

Fig. \ref{fig:gini_m20} shows the Gini--$M_{20}$ diagram for our observational Pan-STARRS sample (left-hand panel), as well as for galaxies from the IllustrisTNG (middle panel) and original Illustris (right-hand panel) simulations. All panels show objects with $\log_{10}(M_{\ast}/\Msun) \approx 9.8$--$11.3$ at $z \approx 0.05$. The bin colours indicate the median concentration index ($C$) of the galaxies in each bin, while the contours contain 68 and 95 per cent of the galaxy population. The orange and blue lines, introduced in \cite{Lotz2008a}, are commonly used to classify $0.2 < z < 0.4$ galaxies as early-type, late-type, or merger candidates, as indicated in the left-hand panel. These lines are used to define the $G$-$M_{20}$ `bulge' and `merger' statistics \citep{Snyder2015a, Snyder2015}.

Evidently, the locus of the Gini--$M_{20}$ diagram in IllustrisTNG is in very good agreement with the one derived from observations. In particular, the median concentration of IllustrisTNG galaxies within the inner contour (containing 68 per cent of the sample) ranges from $C \sim 2.5$ to $C \sim 4$, just like in the observations. On the other hand, the bulk of the galaxy population in the original Illustris population has significantly lower median concentrations ($C \sim 2$--$3$), while the $G$-$M_{20}$ locus lies at lower Gini values (i.e. the pixel flux distribution becomes more homogeneous) and higher $M_{20}$ values (i.e. the brightest 20 per cent of the galaxy's light becomes more spatially extended). This range of $G$ and $M_{20}$ values is roughly consistent with the measurements of \cite{Bignone2017} for $g$-band images of Illustris galaxies at $z=0$. Note, however, that Fig. \ref{fig:gini_m20} is not directly comparable to fig. 2 from \cite{Snyder2015}, which shows galaxies at a higher mass range.

The improved morphological realism of IllustrisTNG galaxies evident from Fig. \ref{fig:gini_m20} is primarily due to two reasons. First, the reparametrization of the galactic winds \citep{Pillepich2018} results in galaxies with roughly correct sizes at the low-mass end, while also producing thinner discs. Second, the new AGN feedback model \citep{Weinberger2017} is able to quench galaxies more efficiently at the high-mass end, allowing dissipationless processes -- in particular dry mergers -- to have an impact on the morphologies of galaxies that are already quenched \citep{Rodriguez-Gomez2017}.

Nevertheless, Fig. \ref{fig:gini_m20} also reveals the existence of IllustrisTNG galaxies with somewhat low $M_{20}$ values (and correspondingly high Gini coefficients and concentrations) compared to observations. This discrepancy primarily affects massive galaxies. While a detailed investigation of this issue is beyond the scope of this paper, we offer two possible explanations. The first one is related to the numerical implementation of the galaxy formation model. When the kinetic mode AGN feedback is active, momentum is imparted in a weighted fashion to a fixed number of neighbouring gas cells (256 in the fiducial TNG100 run). It is possible that this `feedback injection region' is too large, such that even if the overall galaxy becomes quenched, there is not enough quenching at the smallest galactocentric distances, leaving some residual star formation at the centres of massive galaxies. The second possibility involves the natural processes considered to be important in the formation of massive elliptical galaxies, such as dry mergers \citep[e.g.][]{Khochfar2003, Naab2006a, Rodriguez-Gomez2017}. The importance of these mechanisms might be exaggerated in the IllustrisTNG model due to the sharp transition between the two AGN feedback modes. This picture is supported by Fig. \ref{fig:fgas_vs_mstar}, which shows a sudden drop in the gas fractions of galaxies with $M_{\ast} \gtrsim 10^{10.5} \, \Msun$.

Another notable difference between the $G$-$M_{20}$ diagrams of the Pan-STARRS and IllustrisTNG galaxies is that the latter seems to lack merging systems. This is not an indication of missing mergers in IllustrisTNG but is simply a consequence of how the synthetic images were created. By construction, each image contains a single galaxy identified by the \textsc{subfind} substructure finder, effectively biasing our simulated sample against early-stage mergers. This was done for the sake of simplicity and corresponds to the choice made in \cite{Snyder2015}.

Figs \ref{fig:morph_vs_mstar_1} and \ref{fig:morph_vs_mstar_2} show median trends -- as a function of stellar mass -- for an ample selection of morphological, shape and size measurements (Section \ref{sec:optical_morphologies}), as indicated by the $y$-axis label in each panel. The black lines and grey shaded regions indicate the median and the 16th to 84th percentile range (at a fixed stellar mass) for the Pan-STARRS observations. The blue and red lines correspond to the IllustrisTNG and original Illustris simulations, respectively, with different line styles indicating different models: the solid lines represent our fiducial \textsc{skirt} implementation, including full dust modelling and emission lines from star-forming regions, while the dashed and dotted lines correspond to simpler models (without radiative transfer) including effects from an unresolved dust distribution (i.e. \citealt{Bruzual2003} combined with \citealt{Charlot2000}) and without dust (i.e. a `vanilla' \citealt{Bruzual2003} implementation), respectively.

Figs \ref{fig:morph_vs_mstar_1} and \ref{fig:morph_vs_mstar_2} clearly show that the IllustrisTNG simulation is in better overall agreement with observations than the original Illustris simulation, with median trends in all of the morphological measurements lying within the $\sim$1$\sigma$ scatter of the Pan-STARRS data. It is worth noting, however, that the observational data (solid black lines) always display more monotonic trends with stellar mass, which could be caused by uncertainties (usually around $\sim$0.2 dex) in the stellar mass estimates. In IllustrisTNG (blue lines), the transition seen in many panels around $M_{\ast} \approx 5 \times 10^{10} \, \Msun$ is possibly related to the onset of the kinetic mode AGN feedback, which approximately happens for galaxies above that mass \citep{Weinberger2017}. Naturally, we would find smoother trends in the simulations if we randomly perturb the stellar masses of the simulated galaxies by some amount reflecting the uncertainty in the observations.

In agreement with \cite{Snyder2015}, we find that the original Illustris simulation produces qualitatively correct trends for parameters such as the Gini coefficient ($G$), $M_{20}$, and the concentration index ($C$), although galaxies in general tend to be too large and disc-like compared to observations. This is also consistent with the results of \cite{Bottrell2017} for the Illustris simulation. It is also worth noting that the $F(G, M_{20})$ bulge statistic, the concentration index ($C$) and the S\'{e}rsic index ($n$) display very similar trends with stellar mass. In fact, these quantities will be considered to be essentially interchangeable in the following sections. Finally, Figs \ref{fig:morph_vs_mstar_1} and \ref{fig:morph_vs_mstar_2} also show that the inclusion of dust effects has a non-negligible effect on galactic structure, making galaxies at the low-mass end less concentrated near their centres and more spatially extended. 

\subsection{Morphology and galaxy colour}\label{subsec:bulge_statistic_vs_g-i}

\begin{figure*}
	\centerline{
		\vbox{
			\includegraphics[width=17.5cm]{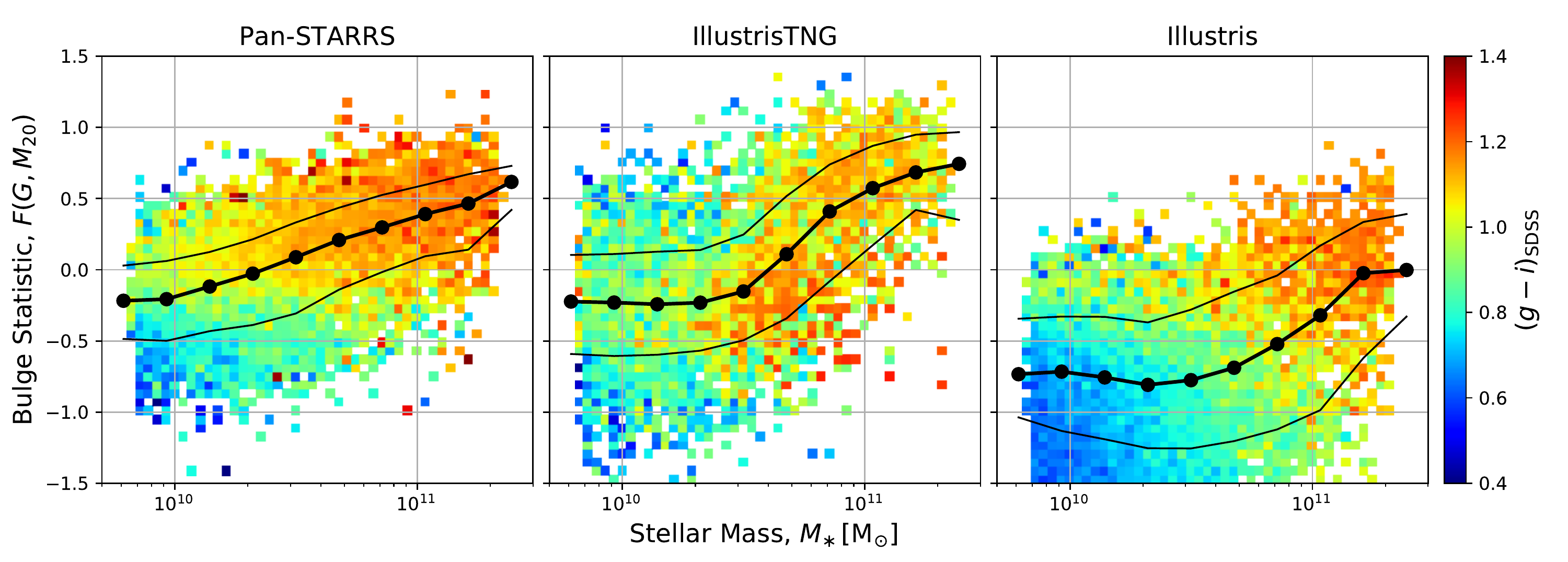}
		}
	}
	\caption{The Gini--$M_{20}$ bulge statistic, $F(G, M_{20})$, as a function of stellar mass. The colour of each 2D bin corresponds to the median $g-i$ colour index of the galaxies that fall into that bin. The left-hand panel shows galaxies observed with Pan-STARRS, while the middle and right-hand panels show galaxies from the IllustrisTNG and Illustris simulations, respectively. In each panel, the thick and thin lines show the median and the 16th to 84th percentile range at a fixed stellar mass. As evidenced by the colour gradients, galaxy morphology and colour are strongly correlated both in observations and in the original Illustris simulation. In the case of IllustrisTNG, however, the colour gradient becomes nearly horizontal, which means that galaxy colour is more strongly correlated with stellar mass than with morphology. The same applies to other morphological indicators, such as concentration ($C$) or S\'{e}rsic index ($n$).}
	\label{fig:bulge_statistic_vs_mstar_by_g-i}
\end{figure*}

\begin{figure*}
  \centering
	\includegraphics[width=17.5cm]{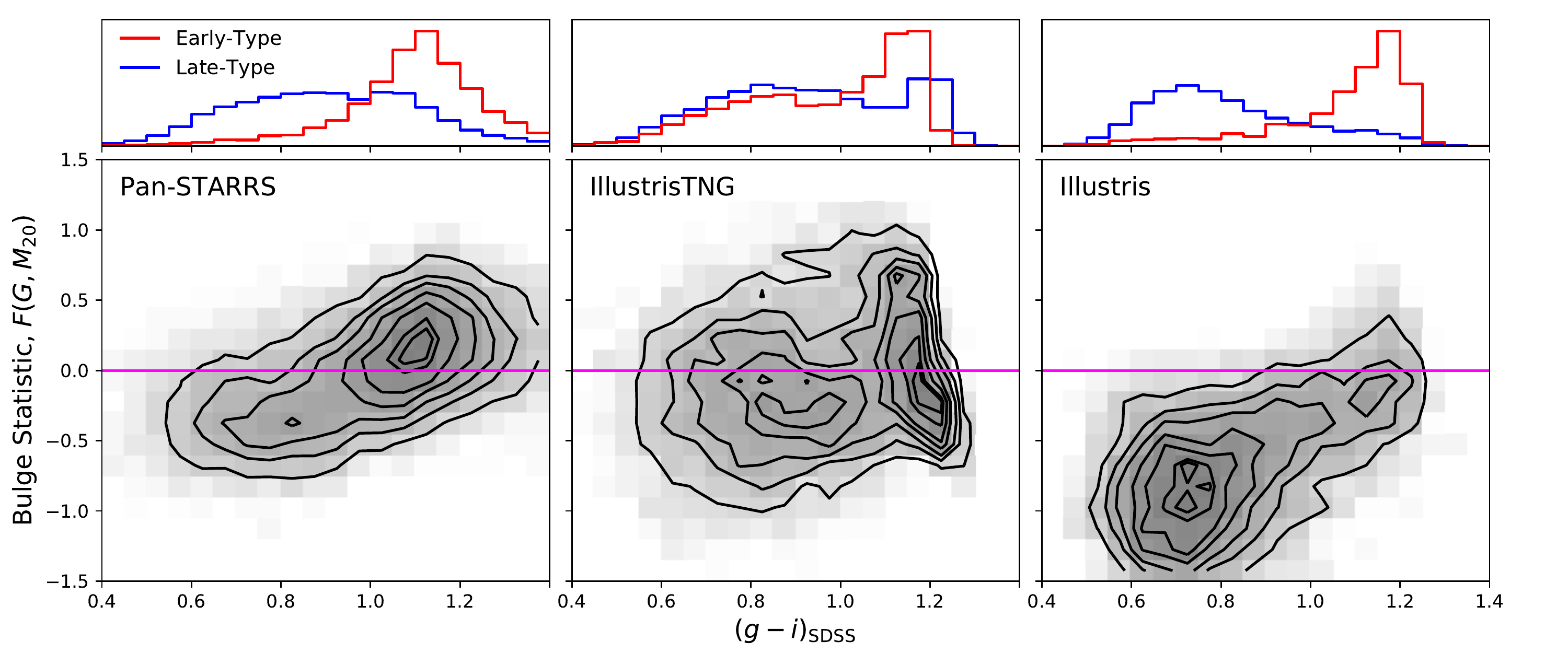}
	\caption{Distribution of the $F(G, M_{20})$ bulge statistic as a function of the $g-i$ colour index for the full galaxy sample ($\log_{10}(M_{\ast}/\Msun) \approx 9.8$--$11.3$). The left-hand panel shows galaxies observed with Pan-STARRS, while the middle and right-hand panels show galaxies from the IllustrisTNG and Illustris simulations, respectively. In the bottom panels, the grey scale and contours show the overall galaxy distribution, while the histograms in the upper panels show the marginal distributions of early-type and late-type galaxies, defined as those with $F(G, M_{20}) \geq 0$ and $F(G, M_{20}) < 0$, respectively. While IllustrisTNG features a clear colour bimodality, it also produces a relatively large proportion of red ($g-i \geq 1$) discs and blue ($g-i < 1$) spheroids relative to observations. We obtain qualitatively similar results if we replace $F(G, M_{20})$ with the concentration index ($C$) or the S\'{e}rsic index ($n$).}
	\label{fig:bulge_statistic_vs_g-i}
\end{figure*}

\begin{figure*}
	\centerline{
		\vbox{
			\includegraphics[width=17.5cm]{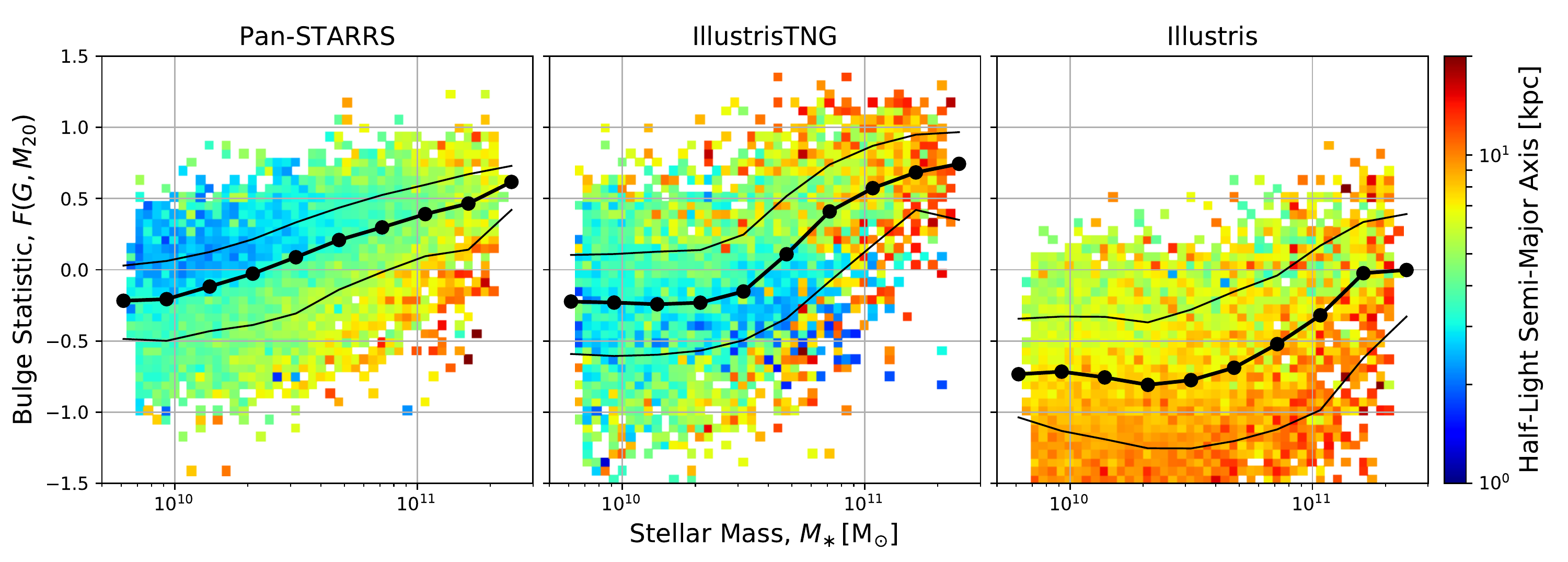}
		}
	}
	\caption{The Gini--$M_{20}$ bulge statistic, $F(G, M_{20})$, as a function of stellar mass. The colour of each 2D bin indicates the median size (specifically, the semi-major axis of an ellipse containing half of the total flux) of the galaxies that fall into that bin. The left-hand panel shows galaxies observed with Pan-STARRS, while the middle and right-hand panels show galaxies from the IllustrisTNG and Illustris simulations, respectively. In each panel, the thick and thin black lines show the median and the 16th to 84th percentile range at a fixed stellar mass. Both in the observations and in the original Illustris simulation, discs tend to be larger than spheroids at any given stellar mass, while the trend is less clear in IllustrisTNG. The same is true if we replace $F(G, M_{20})$ with the concentration index ($C$) or the S\'{e}rsic index ($n$), or if we use S\'{e}rsic half-light radii.}
	\label{fig:bulge_statistic_vs_mstar_by_rhalf}
\end{figure*}

\begin{figure*}
  \centering
	\includegraphics[width=17.5cm]{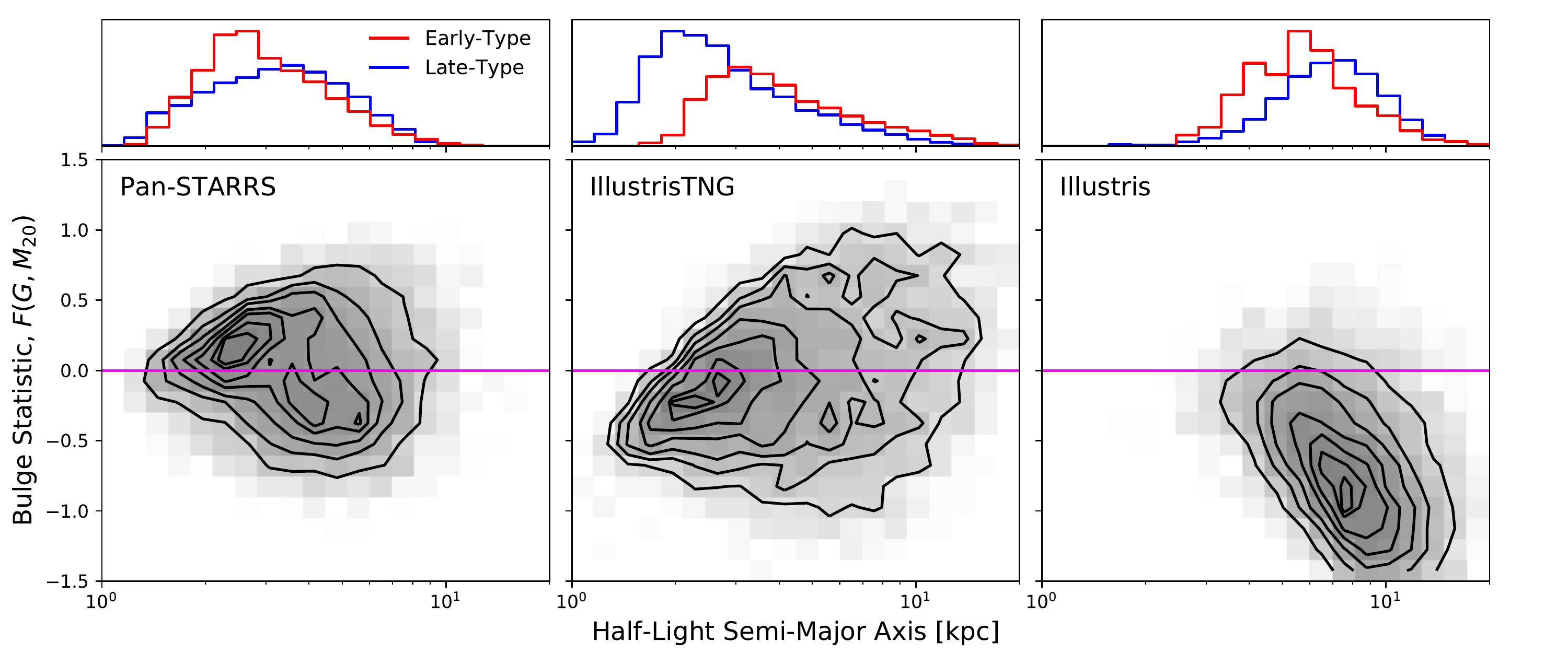}
	\caption{Distribution of the $F(G, M_{20})$ bulge statistic as a function of galaxy size (namely, the semi-major axis of an ellipse containing half of the total flux) for the full galaxy sample ($\log_{10}(M_{\ast}/\Msun) \approx 9.8$--$11.3$). The left-hand panel shows galaxies observed with Pan-STARRS, while the middle and right-hand panels show galaxies from the IllustrisTNG and Illustris simulations, respectively. In the bottom panels, the grey scale and contours show the overall galaxy distribution, while the histograms in the upper panels show the marginal distributions of early-type and late-type galaxies, defined as those with $F(G, M_{20}) \geq 0$ and $F(G, M_{20}) < 0$, respectively. This figure shows that the correlation between galaxy morphology and size in IllustrisTNG is in tension with the observations. We obtain similar results if we replace $F(G, M_{20})$ with the concentration index ($C$), or if we plot the S\'{e}rsic index ($n$) versus the S\'{e}rsic half-light radius.}
	\label{fig:bulge_statistic_vs_rhalf}
\end{figure*}

After demonstrating that the optical morphologies of galaxies in the IllustrisTNG simulation are essentially correct to first order for the stellar masses considered, we proceed to explore the connection between morphology and other galaxy properties, such as colour. Fig. \ref{fig:bulge_statistic_vs_mstar_by_g-i} shows the $F(G, M_{20})$ bulge statistic as a function of stellar mass, where the colour of each bin indicates the SDSS $g-i$ colour index. The thick and thin black lines show the median and the 16th to 84th percentile range of $F(G, M_{20})$ at a fixed stellar mass. For the observational sample, the SDSS colours were obtained by matching to the NSA catalogue as described in Section \ref{subsec:obs_sample}. For the simulation sample, we generated mock SDSS images following the same methodology of Section \ref{sec:synthetic_images} (this time using SDSS broad-band filter curves and a pixel scale of 0.396 arcsec pixel$^{-1}$) and then calculated the ratio between the Petrosian fluxes (integrated over a circular aperture with radius equal to \textit{twice} the Petrosian radius, following the SDSS convention) in the $g$ and $i$ bands.

As already mentioned in the previous section, the overall $F(G, M_{20})$ values of simulated galaxies show better agreement with observations in IllustrisTNG than in Illustris original. Fig. \ref{fig:bulge_statistic_vs_mstar_by_g-i} shows that this is true not only for the median trend, but also for the $1\sigma$ scatter in $F(G, M_{20})$ at a fixed stellar mass, which is about twice as large in the original Illustris simulation than in the observations, while the discrepancy between IllustrisTNG and Pan-STARRS is $\sim$30 per cent.

Fig. \ref{fig:bulge_statistic_vs_mstar_by_g-i} also shows that bulge strength is very strongly correlated with galaxy colour in the observations (left-hand panel), as evidenced by the nearly vertical gradient of the $g-i$ colour index in the $M_{\ast} - F(G, M_{20})$ plane. A strong correlation between morphology and colour at a fixed stellar mass is also apparent for the original Illustris simulation (right-hand panel), despite the fact that it overproduces galaxies with late-type morphologies. Surprisingly, the aforementioned colour gradient is approximately horizontal in the IllustrisTNG simulation (middle panel), which suggests that galaxy morphology and colour are relatively independent from each other in the newer model.

Fig. \ref{fig:bulge_statistic_vs_g-i} further explores the connection between the $F(G, M_{20})$ bulge statistic and the $g-i$ colour index by plotting these quantities directly against each other, as shown by the grey scale and contours in the lower panels. This figure includes the full galaxy sample, with stellar masses $\log_{10}(M_{\ast}/\Msun) \approx 9.8$--$11.3$. The histograms in the upper panels show the relative distribution of galaxy colours for early-type and late-type galaxies, defined as those with $F(G, M_{20}) \geq 0$ and $F(G, M_{20}) < 0$, respectively. While such distributions are clearly different from each other in the observations and in the original Illustris model, they are quite similar in IllustrisTNG. We reach the same qualitative trends if we individually examine narrower stellar mass ranges (specifically, $\log_{10}(M_{\ast}/\Msun) =$ 9.8--10.3, 10.3--10.8, and 10.8--11.3), as well as if we replace $F(G, M_{20})$ with the concentration index ($C$) or the S\'{e}rsic index ($n$). These trends are also qualitatively consistent with the fact that the specific star formation rate (sSFR) and the effective stellar mass surface density $(\Sigma_{\rm e})$ are less strongly correlated in IllustrisTNG than in observations \citep[][their figures 8 and 12]{Habouzit2018}.

The trends described above result in a higher fraction of red discs and blue spheroids relative to observations. Defining a red/blue cut at $g-i = 1$, the percentages of red discs and blue spheroids relative to the total number of discs ($F(G, M_{20}) < 0$) and spheroids ($F(G, M_{20}) \geq 0$) are 35 and 21 per cent for the Pan-STARRS data, compared to 39 and 48 per cent in IllustrisTNG, respectively. If we instead define a threshold between early types and late types at $F(G, M_{20}) = -0.1$, which is perhaps better suited for our low-redshift sample, the aforementioned percentages of red discs and blue spheroids would change to 28 and 24 per cent in the observations, and 37 and 49 per cent in IllustrisTNG. The formation of red discs within the IllustrisTNG framework will be studied in detail in Tacchella et al. (in preparation).

\subsection{Morphology and galaxy size}\label{subsec:bulge_statistic_vs_rhalf}

Fig. \ref{fig:bulge_statistic_vs_mstar_by_rhalf} once again shows the $F(G, M_{20})$ bulge statistic as a function of stellar mass. In this case, however, the 2D bin colours indicate the median galaxy size, parametrized by the semi-major axis of an ellipse containing half of the total luminosity (Section \ref{subsubsec:rhalf}), with redder colours indicating larger galaxies. The different panels show Pan-STARRS observations of galaxies at $z \sim 0.05$ and the corresponding synthetic images from the IllustrisTNG and Illustris simulations. The thick and thin black lines indicate the median and 16th to 84th percentile range (at a fixed stellar mass) of $F(G, M_{20})$, which are identical to those previously shown in Fig. \ref{fig:bulge_statistic_vs_mstar_by_g-i}.

The left-hand panel of Fig. \ref{fig:bulge_statistic_vs_mstar_by_rhalf} shows a strong correlation between the $F(G, M_{20})$ bulge statistic and galaxy size, with disc-like systems being larger, on average, than spheroids in any stellar mass range. This is a well-known observational trend, which is also reproduced by the original Illustris simulation (right-hand panel), despite the fact that Illustris galaxies exhibit systematically large sizes compared to observations. On the other hand, such a well-defined trend is absent in the IllustrisTNG simulation (middle panel).

In Fig. \ref{fig:bulge_statistic_vs_rhalf}, we further explore the relationship between morphology and galaxy size. The bottom panels show the overall distribution of galaxies in the morphology--size plane, while the upper panels show the size distribution of early-type ($F(G, M_{20}) \geq 0$) and late-type ($F(G, M_{20}) < 0$) galaxies separately. We consider the full galaxy sample at $z \sim 0.05$ with stellar masses $\log_{10}(M_{\ast}/\Msun) \approx 9.8$--$11.3$. This figure confirms that late-type galaxies in the observational sample (left-hand panel) and in the original Illustris simulation (right-hand panel) tend to be larger than early-type galaxies, on average, although there is significant overlap between the two populations. Interestingly, in IllustrisTNG the trend is reversed: early-type galaxies tend to be larger than late-type ones.

We find that these qualitative trends continue to hold if we individually examine narrower stellar mass ranges ($\log_{10}(M_{\ast}/\Msun) =$ 9.8--10.3, 10.3--10.8, and 10.8--11.3), as well as if we replace $F(G, M_{20})$ with other structural measurements such as the concentration index ($C$) or the S\'{e}rsic index ($n$), or if we use size measurements derived from S\'{e}rsic profile fitting (Section \ref{subsec:sersic_profiles}) instead of the non-parametric size measurements currently shown.

Recently, \cite{Genel2018} explored galaxy sizes in the IllustrisTNG simulation and found that the sizes of main-sequence galaxies are systematically larger than those of quenched galaxies, in agreement with observations, except at $M_{\ast} \lesssim 10^{9.5} \, \Msun$. We have verified that we can reproduce those results in IllustrisTNG if we replace the $F(G, M_{20})$ bulge statistic in Fig. \ref{fig:bulge_statistic_vs_mstar_by_rhalf} with the sSFR predicted by the simulation. Similarly, replacing $F(G, M_{20})$ with the colour index $g-i$ in Fig. \ref{fig:bulge_statistic_vs_rhalf} would demonstrate that blue galaxies tend to be larger than red ones (within the stellar mass range considered in this paper). In other words, while the morphology--size relation in IllustrisTNG is somewhat problematic, the sSFR-size and colour-size relations are consistent with observations.

\section{Discussion and conclusions}\label{sec:discussion_and_conclusions}

We have made a robust, quantitative comparison between the optical morphologies of galaxies from the state-of-the-art  TNG100 simulation (from the IllustrisTNG suite) and those of galaxies observed with Pan-STARRS. By repeating our analysis on the original Illustris simulation, we have assessed the degree of improvement that the IllustrisTNG galaxy formation model has achieved with respect to its predecessor in the context of image-based galaxy morphology.

We generated $\sim$27,000 synthetic images of $M_{\ast} > 10^{9.5} \, \Msun$ galaxies from the IllustrisTNG and Illustris simulations at $z=0.0485$, designed to match Pan-STARRS observations of galaxies at $z=0.045$--$0.05$. Most synthetic images were created with the radiative transfer code \textsc{skirt}, including the effects of dust attenuation and scattering. The radiative transfer calculations were performed on a reconstruction of the same Voronoi mesh that was used by the \textsc{arepo} code to dynamically evolve the system, which means that the gas density distribution -- used as a proxy for the dust content of the simulated galaxies -- was determined in a self-consistent fashion. Some examples of idealized synthetic images generated with \textsc{skirt} are presented in Fig. \ref{fig:components}.

We analysed both the synthetic images and the real Pan-STARRS images with the newly developed \texttt{statmorph} code, which calculates non-parametric morphological diagnostics such as the Gini--$M_{20}$ \citep{Lotz2004} and CAS \citep{Conselice2003} statistics, and also performs 2D S\'{e}rsic fits. Fig. \ref{fig:morphology_measurements} illustrates some of the various morphological measurements carried out for each galaxy.

We find that the locus of the Gini--$M_{20}$ diagram in IllustrisTNG is consistent with observations (Fig. \ref{fig:gini_m20}), while galaxies from the original Illustris simulation occupy a region of the diagram with lower Gini and higher $M_{20}$ values, which is also characterized by lower concentrations. This shows that bulge-dominated galaxies are under-produced by the original Illustris model, in agreement with \cite{Bottrell2017}.

Overall, the optical morphologies of IllustrisTNG galaxies are in good agreement with Pan-STARRS observations. The median trends (at a fixed stellar mass) for the 18 morphological, shape and size parameters considered in this study are always within the $\sim$1$\sigma$ scatter of the observations in any stellar mass range, which represents a substantial improvement with respect to the original Illustris simulation (Figs \ref{fig:morph_vs_mstar_1} and \ref{fig:morph_vs_mstar_2}). We reiterate that the IllustrisTNG model was not tuned to match image-based morphologies. We also find that the inclusion of attenuation and scattering by a spatially resolved dust distribution has a non-negligible effect on the resulting galaxy morphologies, especially at the low-mass end, making galactic nuclei less concentrated and therefore increasing the galaxies' spatial extent.

However, further inspection shows that the morphology--colour (Figs \ref{fig:bulge_statistic_vs_mstar_by_g-i} and \ref{fig:bulge_statistic_vs_g-i}) and morphology--size (Figs \ref{fig:bulge_statistic_vs_mstar_by_rhalf} and \ref{fig:bulge_statistic_vs_rhalf}) relations in IllustrisTNG are qualitatively inconsistent with observations. Galactic bulge strength, quantified by the $G$-$M_{20}$ bulge statistic, is very weakly correlated with galaxy colour in IllustrisTNG, which results in a somewhat larger fraction of red discs and blue spheroids relative to observations. Furthermore, late-type galaxies in IllustrisTNG tend to be smaller than early types, which is the opposite of what is seen in observations. (Nevertheless, we do find sSFR--size and colour--size relations that are qualitatively consistent with observations, in agreement with \citealt{Genel2018}.) We find similar trends with other structural measurements, such as the concentration index ($C$) or S\'{e}rsic index ($n$).

The lack of a strong connection between galaxy morphology and colour in IllustrisTNG is interesting. Using the Illustris simulation, \cite{Rodriguez-Gomez2017} showed that mergers play a dominant role in shaping galaxy morphology, although only for objects with high stellar masses, which tend to be quenched (this result is reproduced in IllustrisTNG as well). This raises the question of whether mergers also play an important role in quenching galaxies, by triggering powerful starbursts and AGN activity through nuclear gas inflows \citep[e.g.][]{Springel2005a}. Recently, using the `TNG300' simulation from the IllustrisTNG suite (carried out on a larger box than TNG100, measuring $\sim$300 Mpc per side, but at a lower mass resolution), \cite{Weinberger2018} found that galaxy mergers do not play an important role in quenching galaxies within the IllustrisTNG framework.

Previously, however, by running zoom-in simulations of Illustris mergers, \cite{Sparre2016} showed that the mass resolution of the Illustris simulation (which is $\sim$8 times higher than that of TNG300) is not enough to produce starbursts, which were observed in the merger zoom-ins (at a 40 times better mass resolution). In addition, a higher mass resolution could also benefit the current AGN feedback model. By construction, quenching is only achieved in the IllustrisTNG model when the black hole (BH) mass is above $\sim$10$^8 \, \Msun$ \citep{Weinberger2017}. At a higher resolution, a gas-rich merger would also result in more accretion to small galactocentric radii and hence in more BH growth, making the BH more likely to cross the mass threshold needed in the model for it to cause quenching.

Therefore, until we have enough computing power to produce large-scale cosmological simulations at a resolution high enough to capture such starbursts and nuclear gas inflows self-consistently, the implementation of a tunable `merger boost factor' into the galaxy formation model -- i.e. the inclusion of some mechanism that leads to increased quenching via the merger channel -- might be worth exploring. This would also have implications for the morphology--size relation since gas-rich mergers tend to make galaxies smaller (by means of a very dense star-forming core), besides contributing to bulge formation and quenching.

This work highlights the importance of making fair comparisons to observations through forward modelling of simulation data, and by using the same tools to analyse both simulations and observations. An interpretative framework based on hydrodynamic cosmological simulations, such as the one presented here, will be extremely valuable in order to understand galaxy formation in the era of next-generation instruments such as the Large Synoptic Survey Telescope (LSST) and the James Webb Space Telescope (JWST).

\section*{Acknowledgements}

We thank Chris Hayward, Kate Rowlands, Dries Van De Putte, Liza Sazonova, and Mike Fall for useful comments and discussions, as well as Maarten Baes and Peter Camps for making the \textsc{skirt} code public. VRG, JL and GS acknowledge support from the National Science Foundation (NSF) under Grant No. AST-1517559. The POGS catalogue was created with support from NSF grant AST-1412596. This work used the Extreme Science and Engineering Discovery Environment \citep[XSEDE;][]{Towns2014}, which is supported by NSF grant ACI-1548562. The XSEDE allocation TG-AST160043 utilized the Comet and Data Oasis resources provided by the San Diego Supercomputer Center. The IllustrisTNG flagship simulations were run on the HazelHen Cray XC40 supercomputer at the High Performance Computing Center Stuttgart (HLRS) as part of project GCS-ILLU of the Gauss Centre for Supercomputing (GCS). Ancillary and test runs of the project were also run on the compute cluster operated by HITS, on the Stampede supercomputer at TACC/XSEDE (allocation AST140063), at the Hydra and Draco supercomputers at the Max Planck Computing and Data Facility, and on the MIT/Harvard computing facilities supported by FAS and MIT MKI. The original Illustris simulations were run on the Harvard Odyssey and CfA/ITC clusters, the Ranger and Stampede supercomputers at TACC/XSEDE, the Kraken supercomputer at ORNL/XSEDE, the CURIE supercomputer at CEA/France as part of PRACE project RA0844, and the SuperMUC computer at the Leibniz Computing Centre, Germany, as part of project pr85je. The Flatiron Institute is supported by the Simons Foundation.

\bibliographystyle{mnras}

\bibliography{paper}

\end{document}